%% file: main.tex
\documentclass[preprint,review]{elsarticle}

\usepackage{amssymb}

\usepackage[displaymath]{lineno}

\usepackage{siunitx}
\usepackage[version=4]{mhchem}
\usepackage{tabularx}
    \newcolumntype{L}{>{\raggedright\arraybackslash}X}

\newcommand{\um}{\textmu m}
 
\biboptions{numbers,sort&compress }

\journal{Nuclear Instruments and Methods}

\begin{document}
\begin{frontmatter}

\title{The Architected Multi-material Scintillator System: Designs and Modeling }

\author[llnl]{Xianyi Zhang}
\author[llnl]{Jason Philip Brodsky}
\author[llnl]{Elaine Lee}
\author[llnl]{Andrew Neil Mabe}
\author[llnl]{Dominique Porcincula}
\address[llnl]{Lawrence Livermore National Laboratory, Livermore, CA, USA}

\begin{abstract}
We present a new conceptual radiation detector, the Architected Multimaterial Scintillator System, that utilizes a scintillator composed of multiple materials arranged in architected structures to enable new capabilities. By structuring differently-dyed materials, the wavelength of the scintillation light encodes additional information in radiation measurements. These structures can be realized through additive manufacture (3D-printing).
Two classes of this concept are described and modelled using Monte Carlo simulations to evaluate their performance.
The first detector design uses dye microstructures to encode particle tracking information, allowing for directional neutron detection and gamma/neutron discrimination. 
The second design uses a dye gradient to indicate the position of radiation along the gradient. 
The simulation results indicate this new concept in radiation detection can achieve strong performance in a variety of capabilities including particle identification, directionality and spectroscopy measurements, and particle position reconstruction.
\end{abstract}
\end{frontmatter}

\input{narrative/Introduction}

\input{narrative/Simulation}

\input{narrative/PID}

\input{narrative/PRMMSS}

\section{Conclusion}

The simulation studies presented here investigated the performance of emergent advanced particle detection capabilities of conceptual AM scintillators.
These capabilities consist of particle identification, neutron source pointing, neutron spectroscopy, and position resolution, each enabled by a specific architecture of differently-dyed scintillator.
The simulation-based evaluation of these capabilities showed some with great advantages over the existing state of the art and others with conditionally-valuable advantages.

The PID AMSS was shown capable of discriminating between MeV-scale neutron- and gamma-induced interactions.
Although its neutron-gamma distinguishing performance is inferior to PSD, this concept may be valuable in scenarios where PSD is limited by pile-ups from high event rate or in detectors employing low-cost sensors with poor PSD capabilities.

The APID AMSS exhibited demonstrated excellent performance in identifying the direction of a neutron source with high efficiency due to the high efficiency enabled by its sensitivity to recoil angle from single neutron scatters. 
This same track sensitivity also makes the APID AMSS capable of neutron spectroscopy. Further work on this design will investigate the robustness of this performance in an environment with gamma-ray backgrounds and in scenarios where the incident neutron spectrum is unknown.

The PR AMSS showed more precise position resolution than the state-of-the-art using a gradient-dyed scintillator architecture alone. By combining the conventional and gradient position reconstruction approaches, a periodic PR AMSS improved the state-of-art position resolution by an order of magnitude.

The detectors studies in this work assumes the availability of high-performance AM scintillators with light output matching commercially-available scintillators. Research and development is underway by the authors to produce prototype AM scintillators using a number of methods and demonstrate both high light output and realization of the architectures described here.

\section*{Acknowledgements}
This work was performed under the auspices of the U.S. Department of Energy by Lawrence Livermore National Laboratory under Contract DE-AC52-07NA27344.
LLNL-JRNL-820060

\emph{In memory of our dear friend and colleague Andrew Mabe.}


\bibliography{MMSSgroup_natbib}

\end{document}

%% file: narrative/Introduction.tex
\section{Introduction}\label{sec:intro}

Organic scintillators are widely used for particle detection, radiation measurement, and corresponding particle and nuclear physics applications.
They are cost effective, scalable, and can achieve energy resolution on the order of 10\% at \SI{1}{\mega\eV}. 
In addition to measuring energy, scintillation detectors are often equipped to discriminate between recoil particle types and/or identify the position of particle interactions within the scintillator. 
However, these advanced capabilities increase the cost and complexity of scintillation detectors, e.g. requiring specialized pulse-shape discrimination (PSD) scintillator, fast-timing sensors, or subdivision of the detector into individually-instrumented segments. 
These advanced capabilities also have limits determined by statistical uncertainties on the sometimes-small number of photons detected, the emission speed of the scintillator, the detection speed of the photosensors, and the fineness of segmentation.

This paper describes an alternative approach to achieving advanced capabilities in a scintillation detector: the architected multi-material scintillator system (AMSS), a radiation detector using scintillator with a heterogeneous structure consisting of either periodic sub-millimeter segments or a structured blend of two or more scintillator materials.
This multi-material structure creates emergent detection capabilities not present in the base scintillator materials. 
By measuring scintillation properties, such as emission color or timing, an AMSS can determine which material(s) a particle interacted with and so infer where in the structure those interactions occurred. 
Additive manufacturing (AM) technology allows for the production of these multi-material scintillators with precisely controlled geometry as small as tens of \um. 
At this scale, an AMSS can emulate some capabilities of a tracking detector, sensing the track length and angle of recoil particles. 
A different multi-material structure allows for extremely fine position resolution.

\begin{figure}
    \centering
    \includegraphics[width=0.9\columnwidth]{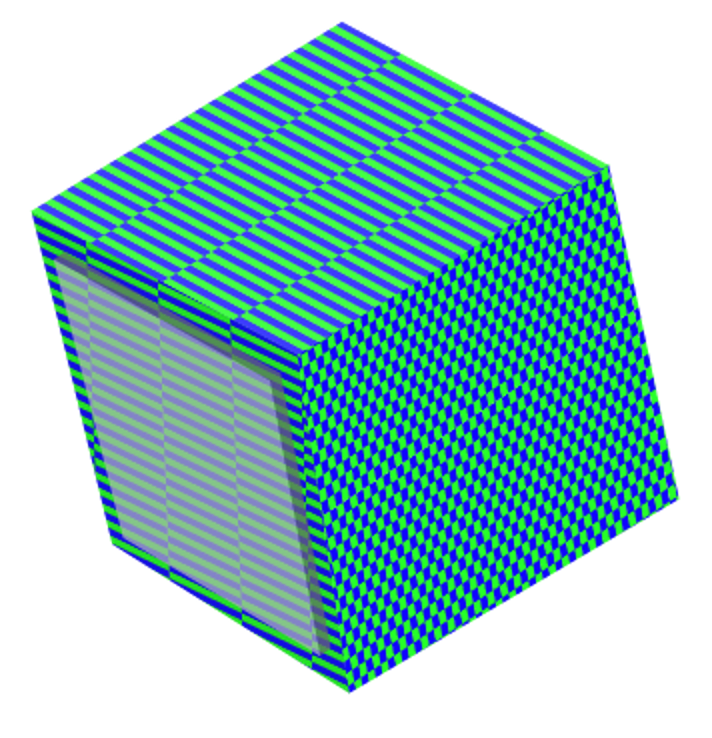}
    \caption{Schematic of a ``checkerboard" manufactured AMSS structure. The green and blue colors of each sub-volume indicate the dyes that shift scintillation light to green and blue wavelengths. The gray box illustrates an arbitrary photosensor. Colors in the figure is to illustrate the structure with differently dyed material. An actual AMSS is expected to be clear due to same refraction indices of the two materials.}
    \label{fig:apid1}
\end{figure}

Although the AMSS concept allows for various structures of a wide range of scintillation properties, in this study, we examined designs for which the color of emitted light is the structured parameter.
Organic scintillators generally incorporate wavelength-shifting dyes inside an ultraviolet-emitting scintillator matrix, in order to shift the light emissions to visible wavelengths efficiently detected by common photosensors.
An AMSS can be composed of differently dyed scintillators arranged in discrete zones within the overall volume, as illustrated in Figure~\ref{fig:apid1}, where the metamaterial is dyed with either blue-emitting or green-emitting dye in a ``3D checkerboard" structure. 
Using filtered light sensors, detection of a single color indicates a recoil particle that remains within a single zone, while detection of both colors indicated a recoil particle that passed through multiple zones. 

Alternatively, blue- and green-emitting dyes can be blended in various proportions to produce a mixture of colors dictated by the dye concentrations. 
If this mixture is varied along a gradient, detecting the proportion of blue and green light produced indicates the location of the particle interaction in that gradient. 
AM enables arbitrary control of the spatial variation of the dye mixture, allowing for bespoke position-sensing structures.

This article describes simulations conducted to evaluate the potential performance of several designs within two conceptual categories of AMSS: particle-track-sensitive and particle-position-sensitive.
In section~\ref{sec:MC}, the Monte-Carlo (MC) simulation methodology is described. 
Section~\ref{sec:PIDMMSS} describes simulations of several track-sensitive detectors and their performance in neutron/gamma discrimination, neutron directionality, and neutron spectroscopy.
Section \ref{sec:PRMMSS} describes simulations of two position-sensitive detectors and their position resolution. 
We conclude by discussing some limitations of the existing simulations and future work to evaluate prototypes of these designs.

%% file: narrative/Simulation.tex
\section{Simulation Methodology}\label{sec:MC}

The simulations reported here were conducted using GEANT4~\cite{agostinelliGeant4SimulationToolkit2003}. 
Neutrons and gammas were generated with spectra of interest to the particular design studied, in the range of 0--10~\si{\mega\eV}. 
Their interactions were tracked in geometries customized for each detector design reported here. 
GEANT4 was also used to simulate optical photons generated by the scintillator and transport to light sensors.

The scintillation production was assigned a nominal yield coefficient of 10000 photons/\si{\mega\eV}. 
Birks' law~\cite{birksScintillationsOrganicCrystals1951} quenching is applied, resulting in a (nonlinear) reduction in light output with respect to quenched energy, especially relevant for the comparison of neutron interactions to gammas.
This simulated Birks' constant is taken from reference~\cite{ashenfelterPROSPECTReactorAntineutrino2019}.

\begin{figure}
    \centering
    \includegraphics[width=0.45\textwidth]{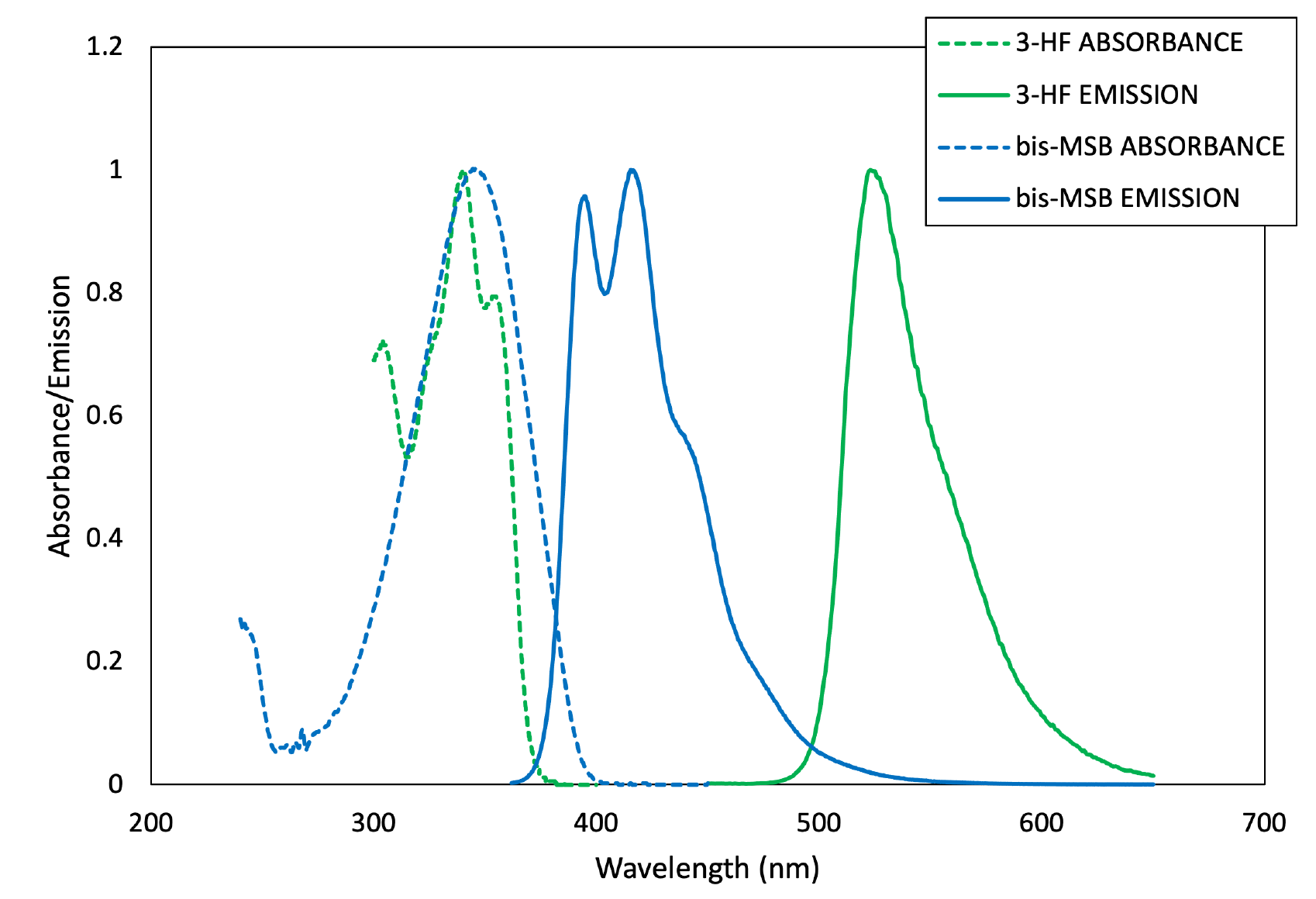}
    \caption{Dye behavior used in the simulation. Shown are the emission and absorption spectra for the blue (bis-MSB) and the green (3-HF) dyes, taken from tests of these dyes performed at LLNL. Spectra are normalized to peak at 1 for comparison. Notably, the blue dye primarily emits in the ``hole'' between the green absorption and emission spectra. As a result, light emitted from blue-dyed zones will not be absorbed or wavelength-shifted when passing through green-dyed zones. The small overlap between the green and blue emission spectra results in a degree of uncertainty in the measured color mixture. }
    \label{fig:dyes}
\end{figure}

As discussed earlier in section~\ref{sec:intro}, a common feature of the designs simulated was the use of two different scintillator dyes. 
In our simulations, light is produced according to emission spectra measured by the authors from bis-MSB (blue-emitting) and 3-hydroxyflavone (3HF, green-emitting). 
The green dye 3HF was chosen because of its exceptionally large Stokes shift with negligible absorption of blue light, as shown in Figure~\ref{fig:dyes}. 
This special property avoids the absorption and wavelength shifting of scintillation light generated from blue-dyed zones.
In these simulations, the same photon yield efficiency was assumed for both scintillators, although measurements suggest 3HF may in reality have a reduced photon yield.

The AMSS detectors are simulated so there is no changes of refractive index at the interfaces between different materials within the scintillator.


In the designs simulated here, the differently colored photons are collected by two photosensors. 
These photosensors, in the simulation, are assumed to have 30\% photon detection efficiencies for both blue and green photons. 
These simulations all have one photosensor coupling to a green-pass filter. 
Some simulations have the other sensor unfiltered while others have that sensor coupling to a blue-pass filter.
The band pass filters are assumed to be perfect step functions and to be index-matched to the photosensor window.

In addition to these common features, each simulation also incorporates a GEANT4 geometry specific to the detector design and multi-material structure.
These geometries 
are provided in the following sections Section~\ref{sec:PIDMMSS} and Section~\ref{sec:PRMMSS}.

%% file: narrative/PID.tex
\section{Track-length-sensitive AMSS}\label{sec:PIDMMSS}

In organic scintillators, radiation interactions with order of \SI{1}{\mega\eV} energy produce recoil tracks with lengths measured on the scale of \SI{10}{\micro\meter}. 
Architected scintillators can be composed of structures in this length scale through additive manufacturing, so they are sensitive to track behavior. 
Organic scintillators have different stopping powers of neutron and electron, and so an AMSS capable of measuring track lengths can be used identify particles. 
This concept, referred to here as a Particle Identification AMSS (PID AMSS) detector encodes track length using different colors of scintillation light. 
A variation on this concept, the Anisotropic AMSS (APID AMSS) has angle-dependent sensitivity to track length, enabling directionality measurements and spectroscopy. 

Track length sensitivity is achieved using a ``checkerboard"-style 3D structure of blue- and green-dyed zones.
These microstructures can be printed 
as small as \SI{50}{\micro\meter} (or smaller using some techniques) which is simultaneously larger than typical neutron-induced proton recoil distances, e.g.~\SI{25}{\micro\meter} at \SI{1}{\mega\eV},
while shorter than gamma-induced electron recoil distances, e.g.~$\sim\SI{144}{\micro\meter}$ at \SI{100}{\kilo\eV}. 
(Due to quenching, \SI{1}{\mega\eV} proton recoils and \SI{100}{\kilo\eV} produce similar amounts of scintillation.)
As a result, a short-track proton recoil is more likely to remain within a single zone and produce a single color scintillation light, while a long-track electron recoil induced by incident gammas is likely to pass through multiple zones and produce both colors of scintillation light. 
If the blue and green zones are cubic, the detector will have an approximately uniform response regardless of recoil direction. 
If the zones are anisotropic, the detector's response is angle dependent.

This PID AMSS structure is different from the conventional segmented detector. 
The latter requires individual instrumentation for each segment while PID AMSS only require two photosensors for the whole volume to distinguish scintillation colors, even with hundreds of millions of zones.


\subsection{Neutron-Gamma Discrimination}\label{sec:PID_n_g}

\begin{figure}[h]
    \centering
    \includegraphics[width=0.45\textwidth]{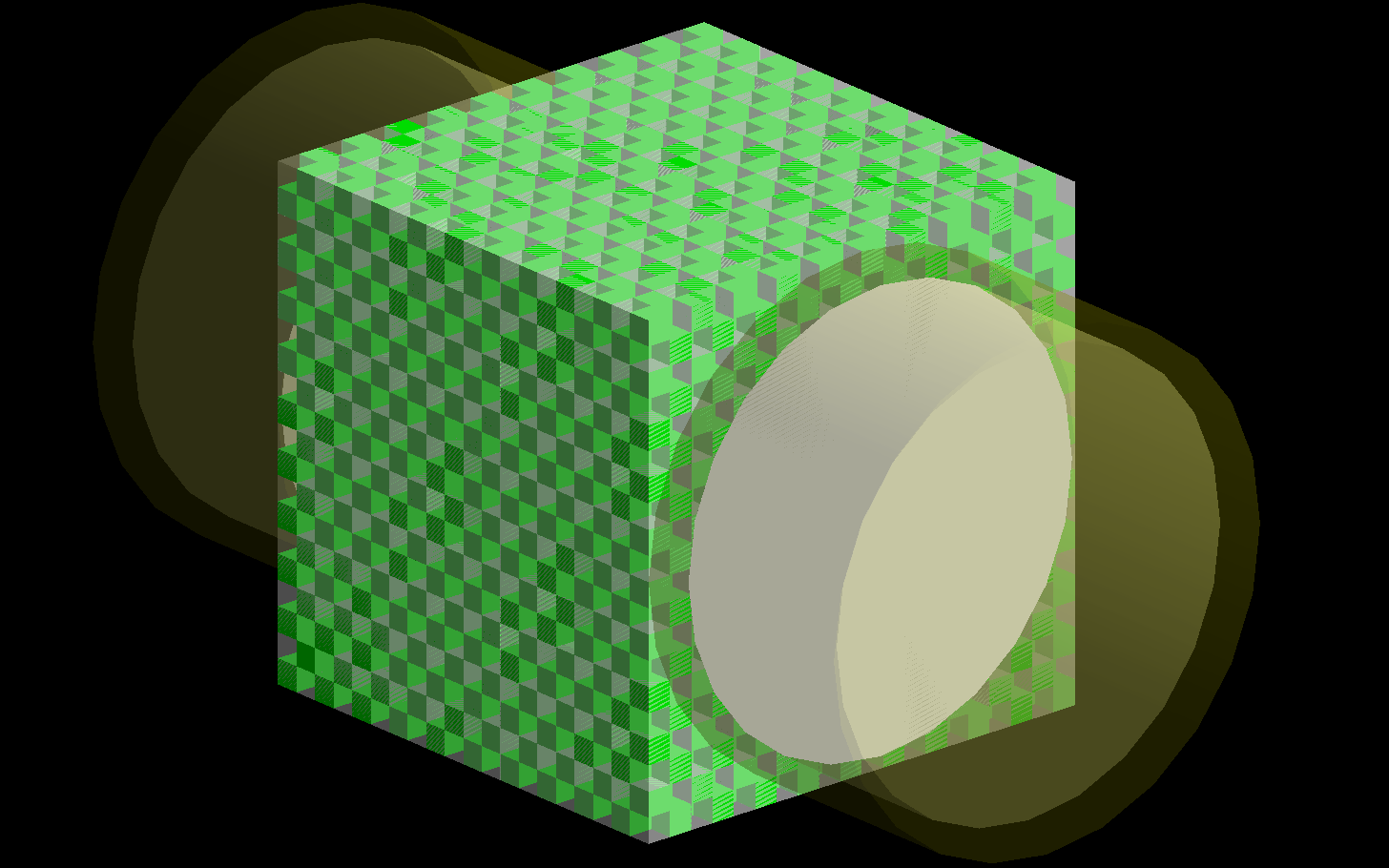}
    \caption{Example of a PID AMSS detector geometry. The cubic target volume is composed of green- and blue-emitting scintillators in a pattern of alternating cubes (blue-emitting scintillator is shown as transparent in this illustration). Two light sensors (yellow) observe the target volume.}
    \label{fig:pidmodule}
\end{figure}

In simulation, the PID AMSS is exposed to neutrons and gammas to evaluate its discrimination ability.
The simulated geometry resembles that shown in Figure~\ref{fig:pidmodule}.
The overall scintillator target is \SI{15x15x15}{\cm} and wrapped in a reflective film except at two ends where glass-enclosed photosensors are attached. 
The scintillator target is subdivided into green- and blue-emitting zones of either $(\SI{100}{\micro\meter})^3$ or $(\SI{200}{\micro\meter})^3$. 
Each photosensor is filtered to accept either only blue or only green photons. 
These zones are much smaller than could be rendered to produce Figure~\ref{fig:pidmodule} and so that render and similar renders elsewhere in the paper illustrate zones with exaggerated size.

\begin{figure*}
    \centering
    \includegraphics[width=0.45\textwidth]{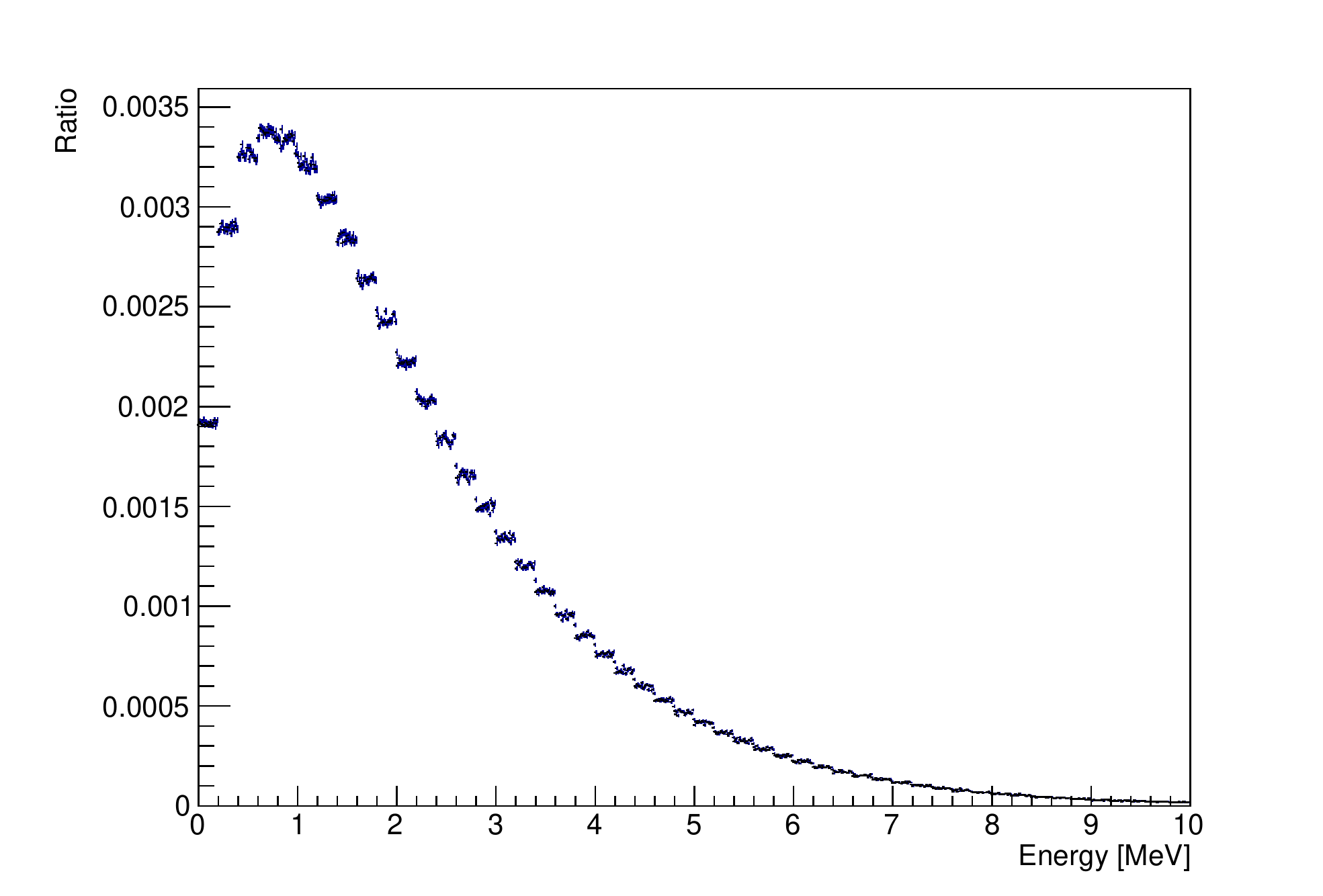}
    \includegraphics[width=0.45\textwidth]{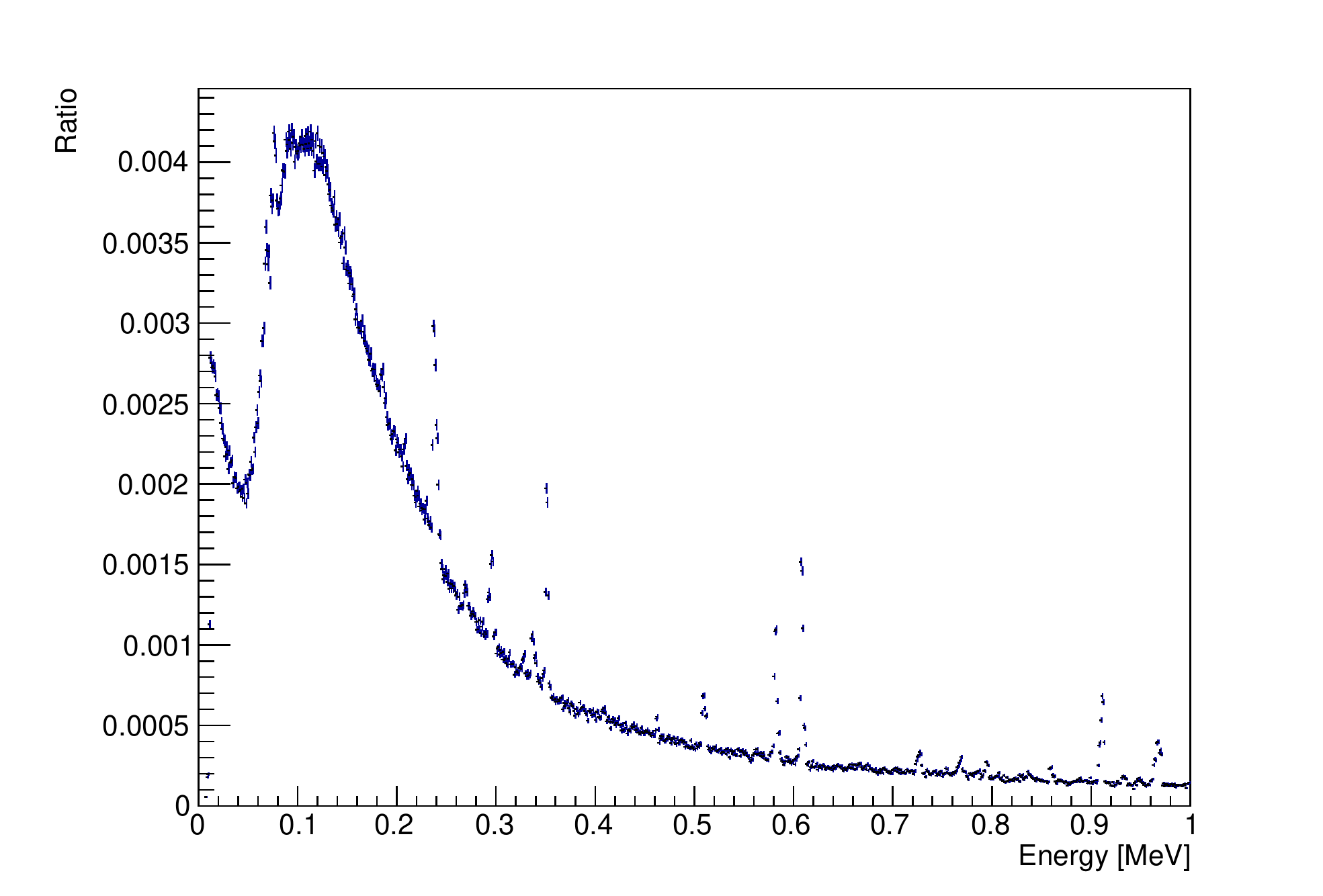}
    \caption{The spectrum of the simulated primary particles. 
    (Left) The Watt spectrum of the neutrons generated.
    (Right) The ambient gamma spectrum generated, based on earth-surface gamma background data.
}
    \label{fig:prim_spec}
\end{figure*}

Neutron events following the Watt neutron energy spectrum~\cite{steinbergerLowStatisticsImagingWeaponsGrade2019} based on \ce{^{252}Cf} were generated from a point source \SI{2}{\m} away from the center of the detector. 
This source produces $1\times10^6$ neutrons per second, similar to a typical $^{252}$Cf source.
Ambient gamma-ray backgrounds were generated uniformly in the detector by approximating the incident gamma spectrum using measurements of gamma interactions in a high-purity germanium detector.
Figure~\ref{fig:prim_spec} shows the energy spectra of the generated neutrons and gammas in this simulation. Because of the quenching effect in organic scintillator, the apparent (``electron equivalent'') energy of 1~MeV proton recoil is close to 0.1~MeV.



The track lengths difference appears in a PID AMSS detector as a difference in the ``color fraction'', i.e.\ ratio of detected photons with two colors.
Color fraction is defined as the fraction of number of photons of the more-detected color to the total number of detected photons, i.e.:
\begin{linenomath}
\begin{align}
    \frac{max(blue,green)}{blue+green}.
\end{align}
\end{linenomath}

\begin{figure}
    \centering
    \includegraphics[width=0.45\textwidth]{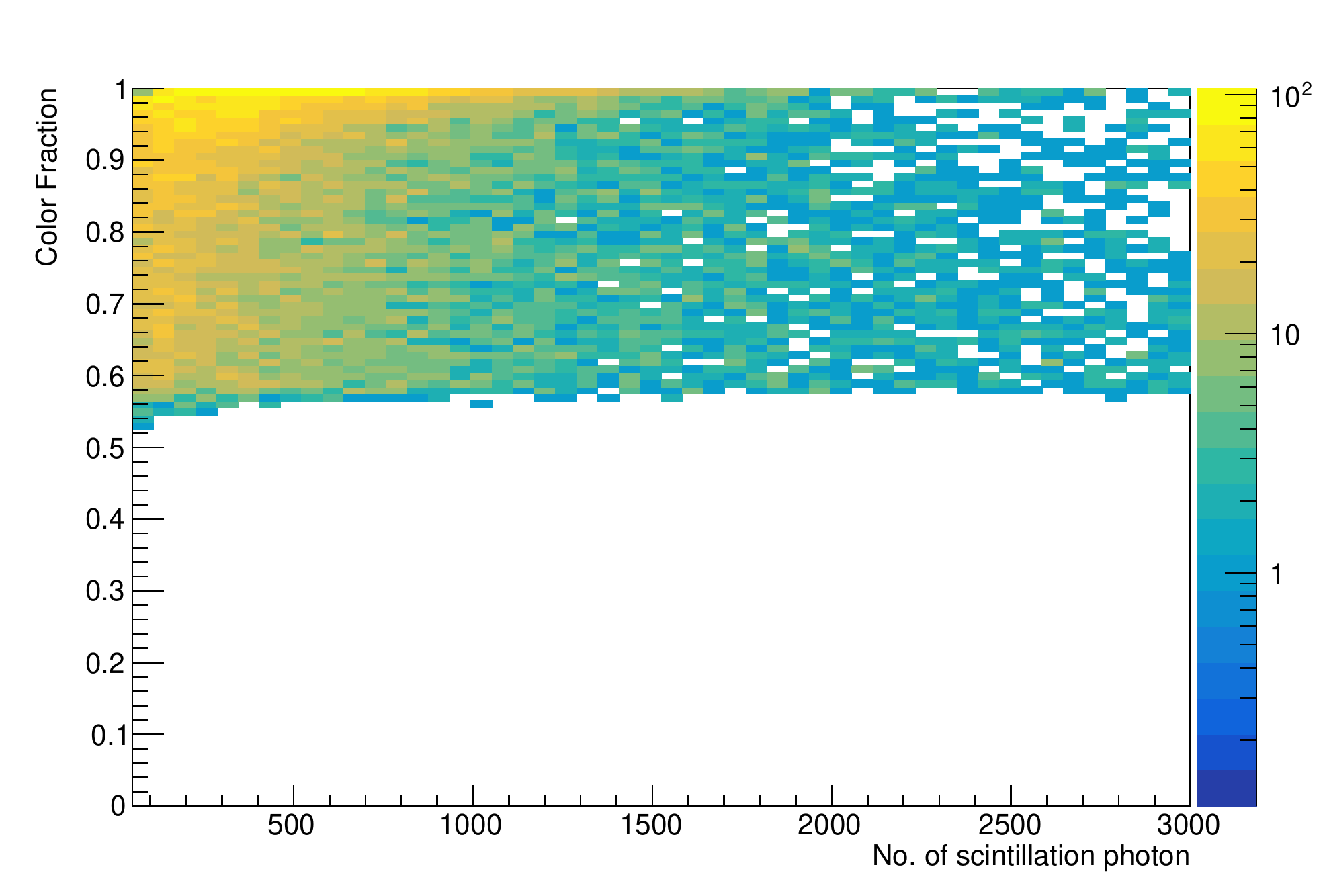}
    \includegraphics[width=0.45\textwidth]{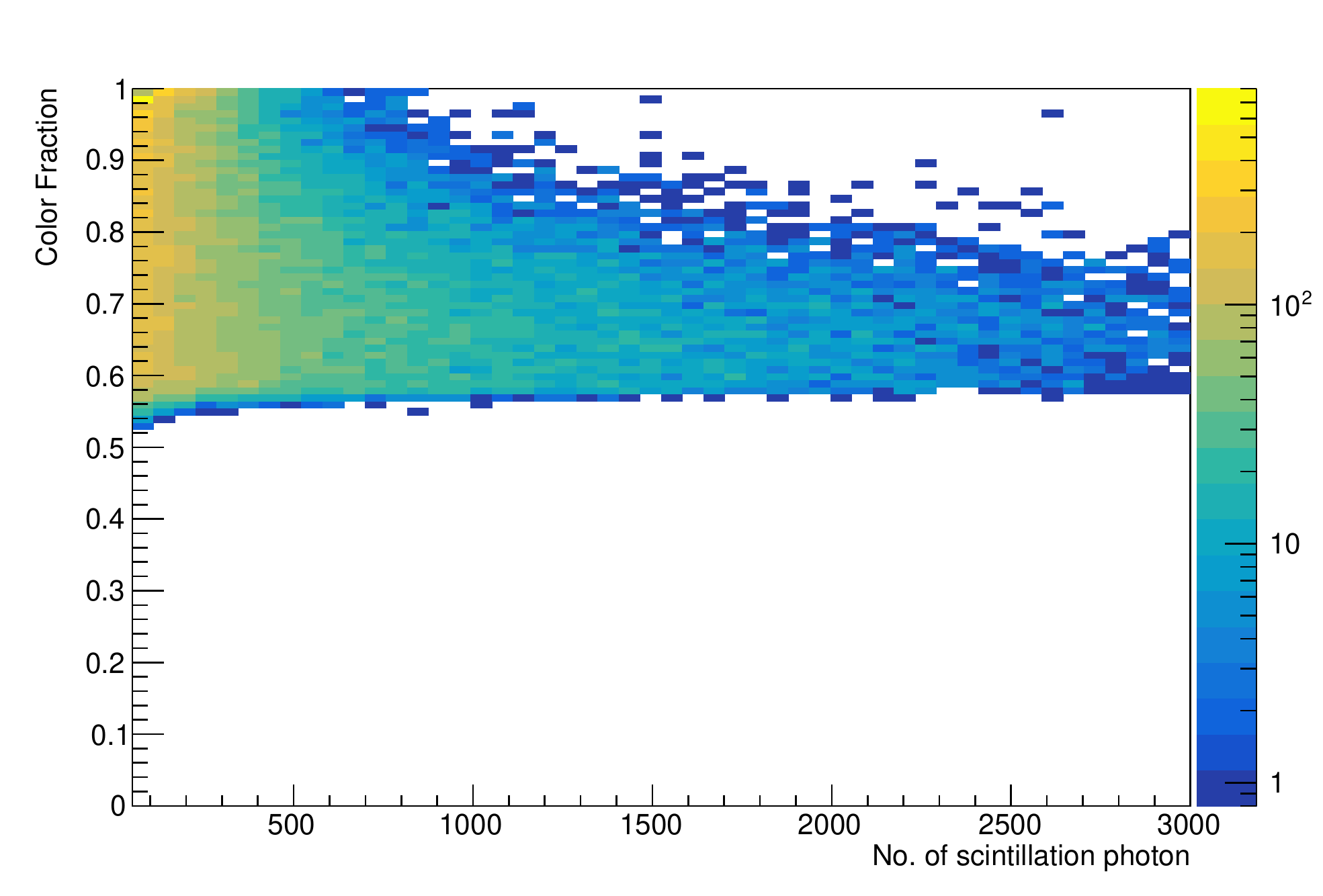}\\
    \caption{Distributions of the color fractions of neutron-proton recoils (top) and gamma-induced electron recoils (bottom) versus the total number of scintillation photons. 
    As particles energy increases, the total number of scintillation photon increase and the gamma and neutron color fraction distributions become more distinct due to the difference in track lengths.}
    \label{fig:color_frac}
\end{figure}

The color fractions produced from the neutron and gamma events in the simulated PID AMSS are shown in Figure~\ref{fig:color_frac}, plotted against the total number of detected scintillation photons.
As expected, the color fraction of neutron induced events is typically 100\%, with a tail towards lower values caused by neutrons that scatter close to a zone boundary allowing the proton recoil to cross between zones.  
This tail is more significant at higher recoil energies where proton recoils travel further to generate scintillation with mixed colors.
At low energy, the color fraction distribution of gamma events is similar to neutrons, as these low-energy electron recoils do not travel far enough to reliably pass between zones. 
With increased gamma recoil energy, the color fractions trend closer to 50\%, becoming clearly distinct from the neutron distribution, as a result of longer tracks crossing zones.

To quantify the performance of neutron gamma discrimination, a tunable cut was implemented using a Support Vector machine Classifier (SVC).
A two-dimensional SVC dependent on color fraction and number of photons was developed using the Python machine learning package \texttt{Scikit-learn} \cite{pedregosaScikitlearnMachineLearning2011} to classify events as neutrons or gammas.

Using a Radial Basis  Function (RBF) kernel, the SVC projects the data from the original dimensions into an infinite dimensional space and then seeks the optimal hyperplane that divides the classes of true particle identity. 
In the original 2D plane of color fraction versus detected photons, this procedure is equivalent to finding a smooth curve that best separates the neutrons from the gammas.  
Varying the distance from the separating hyperplane used to cut events allows evaluation of the neutron acceptance as a function of gamma exclusion, i.e.\ production of a Receiver Operating Characteristic (ROC) curve.

The SVC was trained on 54k simulated gammas and 27k neutrons. It was tested, and a ROC curve produced, using an separate set of 6k gammas and 3k neutrons.

\begin{figure}
    \centering
    \includegraphics[width=0.45\textwidth]{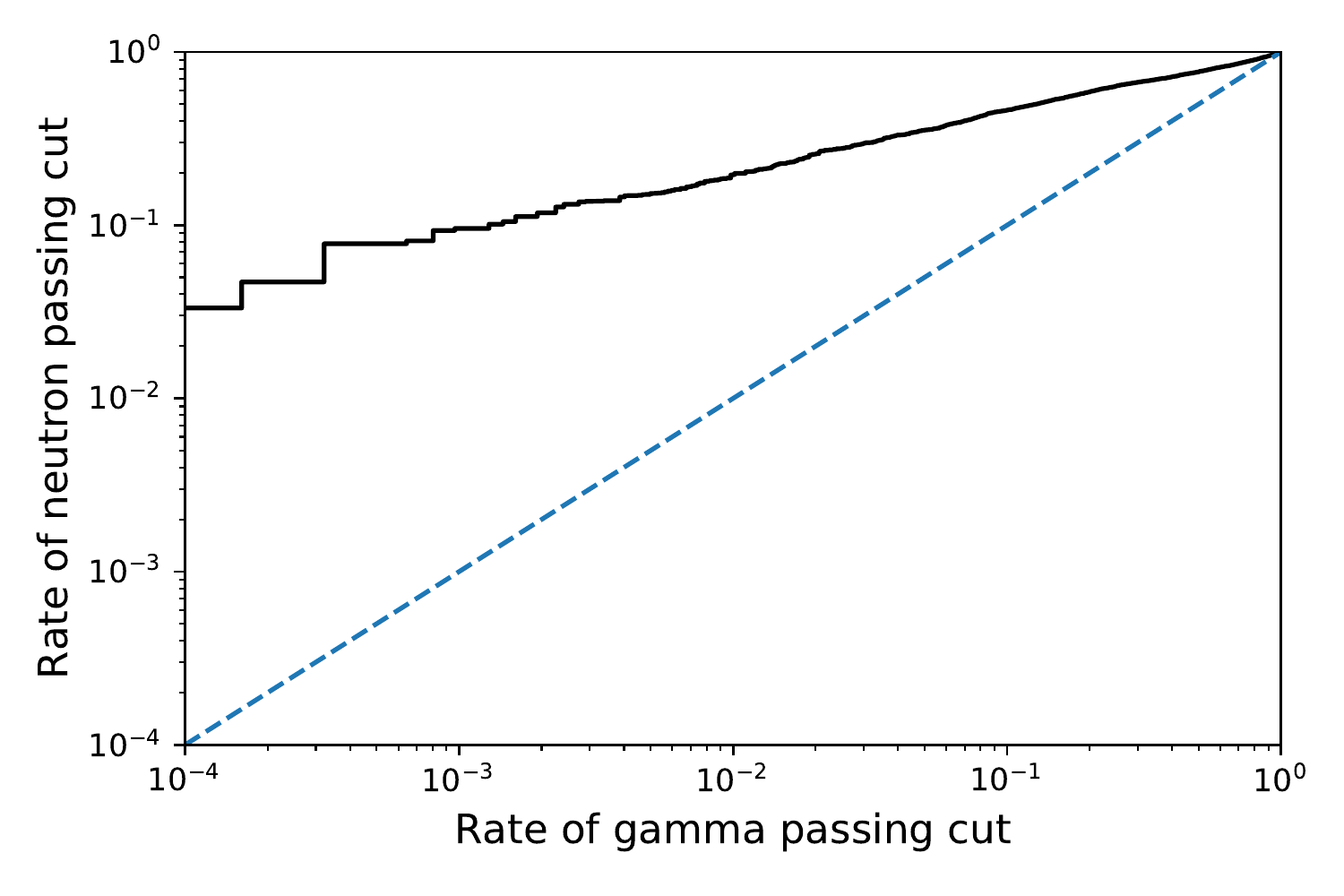}
    \caption{The ROC curve (solid black) of a simulated PID AMSS detector with \SI{200}{\micro\m} zones. The dashed blue line is shown to illustrate equal rates of exclusion.}
    \label{fig:PIDROC}
\end{figure}

The training set includes 90\% of the 60000 simulated gammas and 30000 neutrons. The resulting ROC curve is shown in Figure~\ref{fig:PIDROC}. Table~\ref{tab:pid_results} shows the results for several key points on that curve.

\begin{table}
\begin{center}
\begin{tabular}{ r l} 
\hline
Gamma rejection (\%) & Neutron efficiency (\%) \\
\hline\hline
98.7 & 21 \\
99 & 20 \\
99.9 & 10 \\
\hline
\end{tabular}
\end{center}
    \caption{Gamma rejection and neutron efficiency for PID discrimination. Higher degrees of gamma rejection necessitate a stricter cut that reduces neutron efficiency. Efficiency and rejection are calculated as a fraction of the total Watt spectrum or ambient gamma spectrum, without any energy threshold.}
    \label{tab:pid_results}
\end{table}

\subsection{Neutron-Gamma Discrimination Augmented with Pulse Shape Discrimination}

Neutron-gamma discrimination with PID AMSS operates via a mechanism independent from pulse shape discrimination (PSD) and so the two approaches can complement each other to increase discrimination power.
The PID AMSS places requirements on scintillator dyes, and so is likely incompatible with the dye formulations used for commercial plastic PSD scintillators. 
However, a PID+PSD AMSS may be possible using a scintillator whose PSD properties do not conflict, e.g.\ because they arise from the scintillator matrix rather than the dyes.

In a PID+PSD AMSS detector, the measurement of each radiation event would yield a total photon count, a signal pulse shape, and a color fraction.
These three variables can be analyzed simultaneously for a more definitive discrimination between neutrons and gammas. 

\begin{figure}
    \centering
    \includegraphics[width=0.45\textwidth]{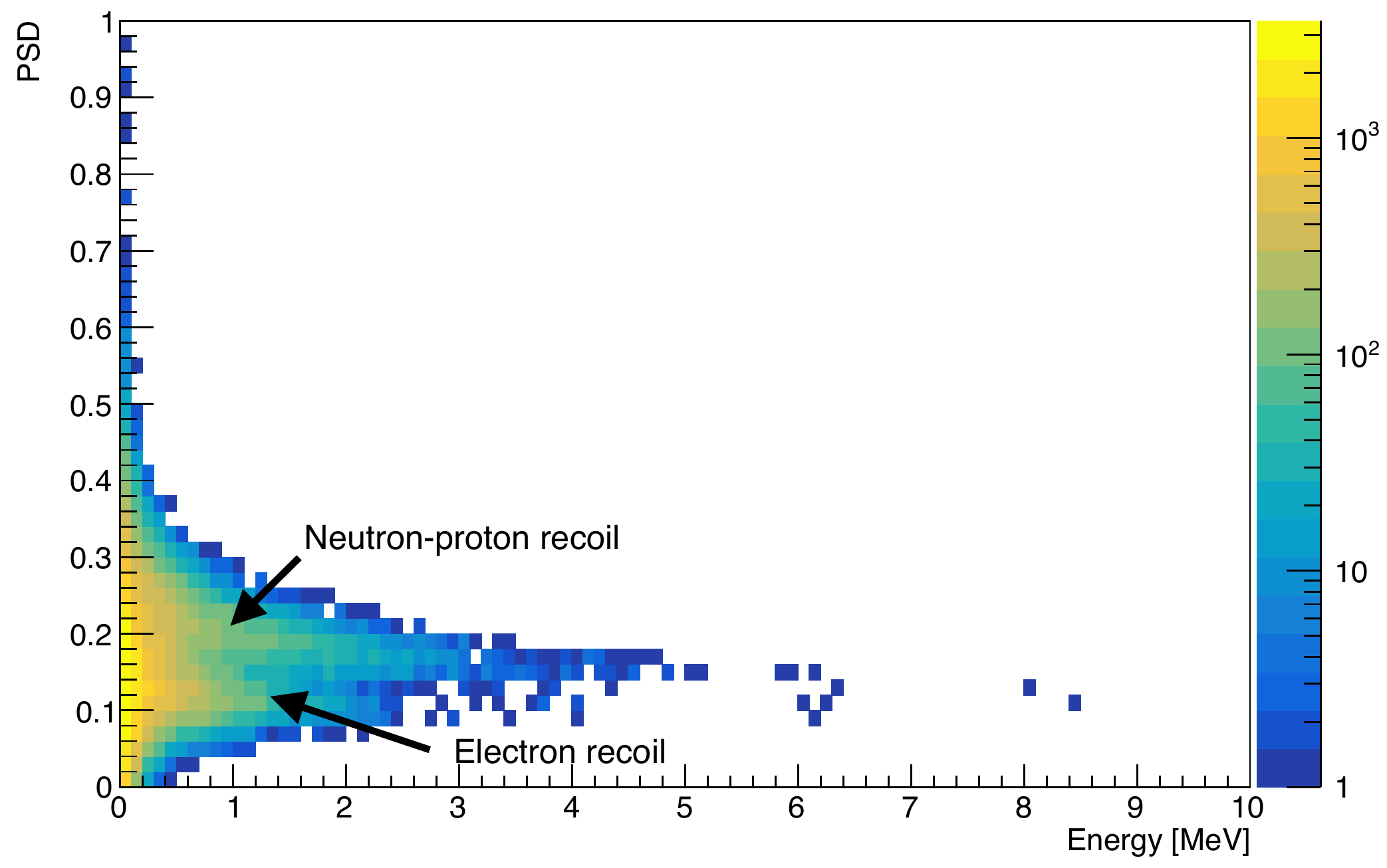}
    \caption{The PSD and energy distribution of simulated gamma and neutron events. The PSD and energy values are both smeared with respect to the reduction of detected photons caused by color filtering.
    }
    \label{fig:PSD}
\end{figure}

\begin{figure}
    \centering
    \includegraphics[width=0.45\textwidth]{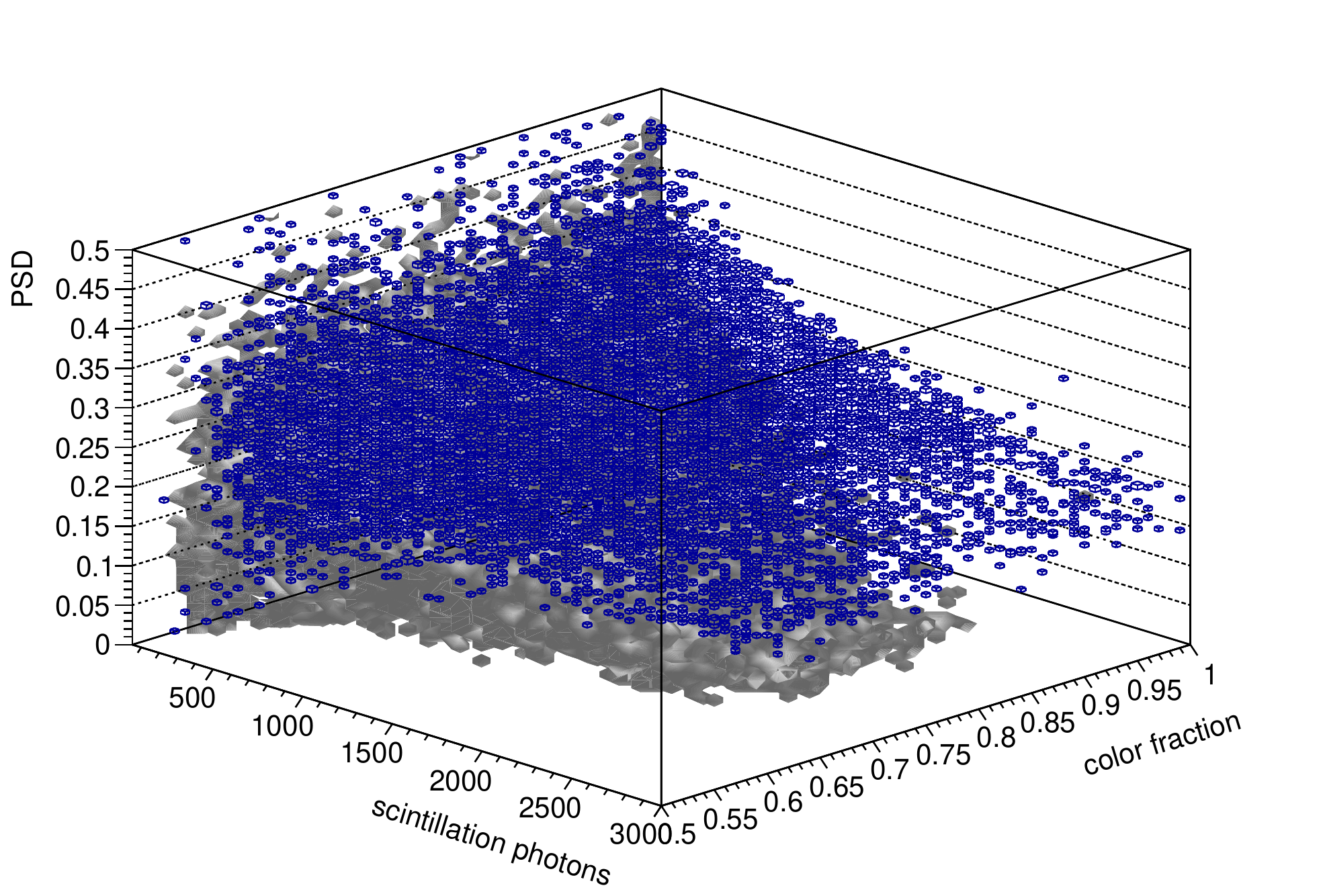}
    \caption{In 3D distribution in the color fraction, PID, and scintillation photons parameter space of neutron (blue) and gamma events (gray).
    The color fraction contributed by PID AMSS is independent from the PSD distribution, allowing for enhanced neutron-gamma discrimination. 
    }
    \label{fig:3D_distribution}
\end{figure}

The events simulated in section~\ref{sec:PID_n_g} were assigned a simulated PSD value drawn from neutron/gamma PSD distributions measured by the the PROSPECT experiment~\cite{ashenfelterPROSPECTReactorAntineutrino2019} with a  PSD-capable scintillator with typical ``as-deployed'' performance~\cite{ashenfelterLithiumloadedLiquidScintillator2019}.
This assignment of PSD values incorporates the statistical variation associated with photon statistics, including the 50\% reduction in photon statistics caused by green/blue color filtration in the AMSS.
Figure~\ref{fig:PSD} shows the assigned PSD and energy distribution of the detected events. Figure~\ref{fig:3D_distribution} extends the events distribution to three dimensions with the addition of a color fraction axis.

\begin{figure}
    \centering
    \includegraphics[width=0.45\textwidth]{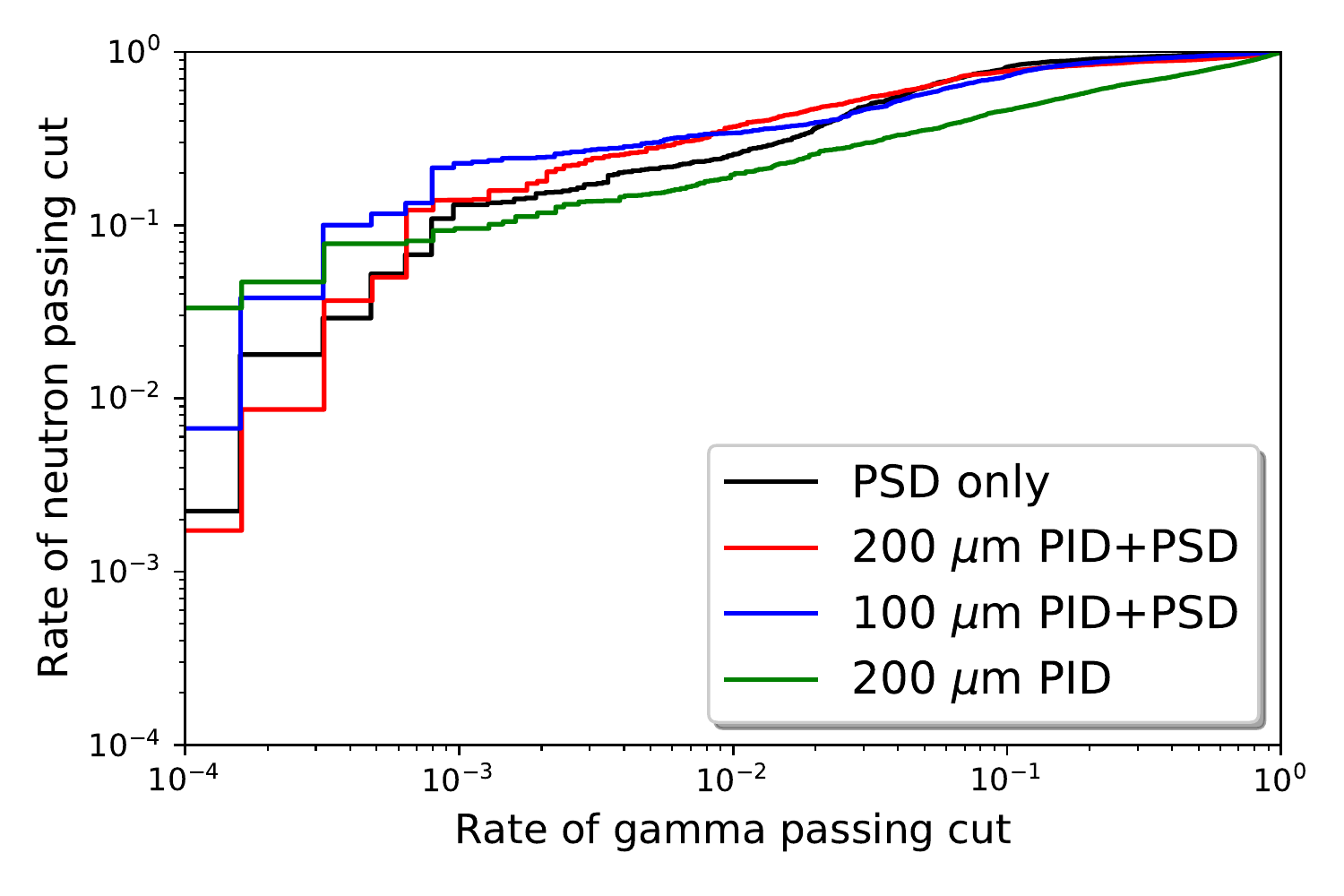}
    \caption{The ROC curves show particle discrimination with PSD scintillator only, PID AMSS only, and PSD with improvement from AMSS.
    Neutron-to-gamma ratio enhanced is achieved even with significantly reduced energy and PSD resolution.
    Comparisons between two PID+PSD AMSS detectors with different sub-volume sizes indicate the discrimination increases with smaller sub-volume.
    }
    \label{fig:PSDROC}
\end{figure}

The RBF SVC method was extended to three dimensions and used to evaluate the effectiveness of adding AMSS structure to a PSD-capable scintillator. 
A major advantage to the SVC approach is the ease of replicating the analysis with either two (PSD-only, PID-only) or three (PID+PSD) input dimensions.
Figure~\ref{fig:PSDROC} and its summary in Table~\ref{tab:pidpsd_results} show that the additional color fraction information from PID AMSS structure enables significantly improved neutron-gamma discrimination. 

Neutron-gamma discrimination using PSD has its worst performance at lower energy because the smaller amount of detected photons reduces the resolution of pulse shape. 
As a result, PID best complements PSD using a smaller zone size that is more sensitive to shorter tracks, as seen in  Figure~\ref{fig:PSDROC}'s comparison of \num{200} and \SI{100}{\micro\m} zones. This is shown in Figure~\ref{fig:PSDROC} and its summary in Table~\ref{tab:pidpsd_results}. 
At 99\% gamma rejection, PSD performs better than PID alone (25\% vs 20\% neutron detection efficiency), while PID+PSD is better than either technology by itself (37\%).

\begin{table*}[ht]
    \centering
    \begin{tabular}{r p{30mm} p{30mm} p{30mm}}
    \hline
    $\gamma$ rejection & $n$ efficiency PSD only & $n$ efficiency PID only & $n$ efficiency (\SI{200}{\micro\m}) PID+PSD \\
        \hline\hline
        98.7\% & 28\%    & 21\% & 40\% \\
        99\% &   25\%    & 20\% & 37\% \\
        99.9\% & 13\% & 10\% & 14\% \\
    \hline
    \end{tabular}
    \caption{Gamma rejection and neutron efficiency for PID discrimination. Higher degrees of gamma rejection necessitate a stricter cut that reduces neutron efficiency. Efficiency and rejection are calculated as a fraction of the total Watt spectrum or ambient gamma spectrum, without any energy threshold.}
    \label{tab:pidpsd_results}
\end{table*}


For many applications, the advantages in discrimination make PID+PSD a compelling option, even at the cost of worsened energy resolution due to light loss in the color filters. 
The PID AMSS alone is conditionally applicable for measurements where PSD is impractical, either due to PSD's requirement for fast-timing sensors or due to high event rates obscuring pulse shapes with pileup events.


\subsection{Anisotropic AMSS for Neutron Source Directionality}\label{sec:APID}

The Anisotropic PID (APID) AMSS uses the alternating-color ``checkboard'' structures of the PID AMSS, except the sub-volumes are longitudinal cuboids, as shown in Figure~\ref{fig:apid1}.
Similar to the PID design, the APID AMSS indicates track length by scintillating with one or two colors, but the anisotropy introduces directional dependence to the track length sensitivity.
The length of each sub-volume in an APID AMSS detector is several times greater than typical track length of a neutron-proton recoil in MeV scale, while the shorter dimensions are sufficient for the recoil to travel across zones.
This directional sensitivity enables the APID AMSS to infer the direction of a neutron source from a sample of proton recoils.

\begin{figure*}
    \centering
    \includegraphics[width=.9\textwidth]{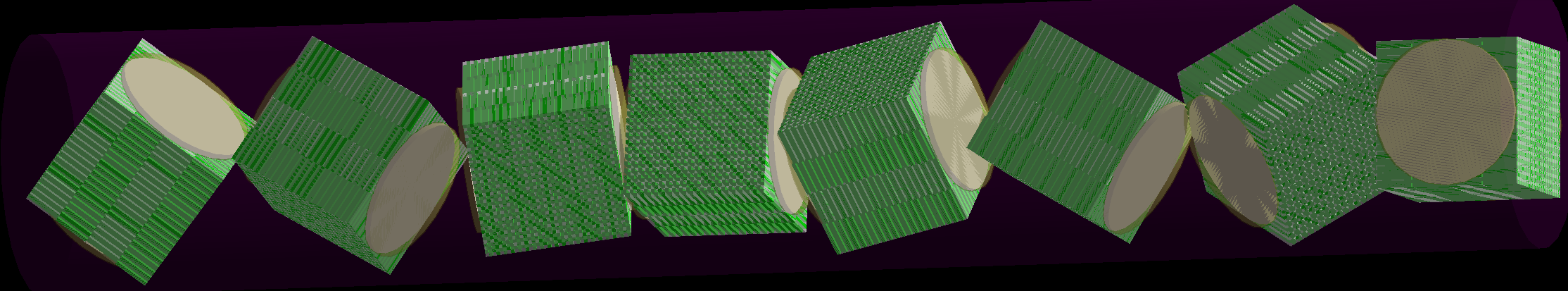}
    \caption{A detector made of 8 APID AMSS modules, each 10~cm $\times$ 10~cm $\times$ 10~cm. Each module has the long axis of the zone structure pointed in a different direction. A fully-realized version of this detector could arrange the modules in a more compact form.}
    \label{fig:apid8}
\end{figure*}

A single APID AMSS detector has a single orientation for the long axis of its zones and is only sensitive to the relative direction between proton recoils and that orientation.
To achieve measurement of the absolute direction of the neutron source in $4\pi$, a detector array was simulated consisting of eight APID AMSS modules, identical except with respect to structure orientation.
The array is illustrated in Figure~\ref{fig:apid8}.
Each module's scintillator target is a \SI{10x10x10}{\cm} cube consisting of \SI{50x50x500}{\micro\m} alternatingly-dyed zones. 
Similar to the PID AMSS detector, each module is read out by two photosensors. 
This simulation explored the requirements on sensor filtration by using one unfiltered sensor and one green-pass-filtered sensor per module. 

To determine the dependence of the array's signals on source direction,  neutrons were simulated from locations uniformly distributed on the surface of a sphere with a radius of 10~m centered on the detector. 
Neutrons were assumed to emit isotropically, but to reduce simulation time only neutrons whose initial direction intersected with the detector array were generated. 
Neutrons were emitted from a Watt spectrum resembling \ce{^{252}Cf}.
Gamma backgrounds, including both ambient gammas and gamma emissions from the $^{252}$Cf reaction were not simulated. 
The potential impact of gamma backgrounds is discussed at the end of this section. 

Any neutrons that generate signals in two or more different modules were excluded from the analysis, because multi-scattering neutrons have a more complex relationship between the observed photon signals and the location of the source.
This is a significant difference between the APID AMSS approach and other approaches to measure neutron source directionality based on kinematic reconstruction (i.e.\ scatter cameras): an APID AMSS neutron detector requires only singly-scattering neutrons, while a scatter camera works only with neutrons that scatter twice. 
About 90\% of simulated neutrons interacting with the array were singly-scattering.
In addition, an energy threshold is applied to exclude events that produce <200 photons.
These cuts applied left the simulated detector array with a 12.6\% neutron detection efficiency among all neutrons that pass through the scintillator. 
This 12.6\% detection efficiency is significantly higher than projected for neutron scatter cameras, which have neutron detection efficiency between 0.1--1\%~\cite{weinfurtherModelbasedDesignEvaluation2018} due to requiring two above-threshold scatters for each neutron. 
With these two cuts, $10^6$ neutrons were used to train the analysis model.

For each neutron, the detector recorded three observable variables: the total number of photons detected, the fraction of green photons to all photons detected, and the ID of the detecting module. 
These observables were related to the two analysis target variables, $\phi$ and $\theta$, the angular coordinates of the true origin of the neutron.
This MC data set was loaded into a five-dimensional Kernel Density Estimator (KDE) to describe the mapping between the three observable and the two target variables. 
This KDE stores the likelihood of a neutron occurring with a specific set of five values based on the distribution of events observed in the training data set. 
The KDE was implemented using the \texttt{Scikit-Learn} Python library.
This approach avoids a functional description of the relationship between measured color(s), proton recoil energy and direction, and source direction, instead relying on the mapping inferred from the large training set.

\begin{figure}
    \centering
    \includegraphics[trim=0 0 450 440,clip,width=.49\textwidth]{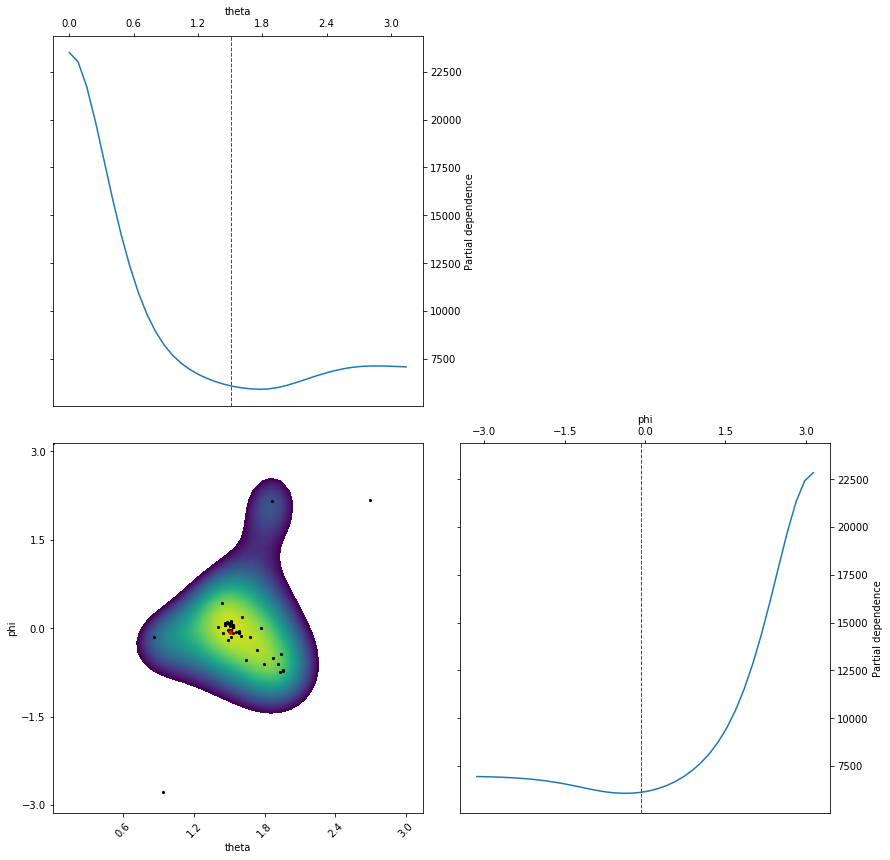}
    \caption{The operation of the Gaussian process minimizer searching for the maximum likelihood using a sample of 1000 neutrons. The minimizer samples points (black) to derive a Gaussian process describing the likelihood function (green/blue contour). This inferred likelihood function helps the minimize choose where to search next to find the maximum likelihood (red), which ends up close to the true value of $(\pi/2,0)$.}
    \label{fig:optimizer}
\end{figure}

The power of this mapping between observables and targets was tested using a simulation set of neutrons from a specific location, $(\theta,\phi) = (\pi/2, 0)$. 
The KDE was queried to return the maximum-likelihood direction given a set of observable values. 
A single neutron is generally insufficient to constrain the source direction, and so ensembles of neutrons of various sizes were selected from the test set, and the KDE was queried for the joint maximum-likelihood direction of the ensemble. 
Searching the KDE for the maximum likelihood direction was organized using the \texttt{Scikit-Optimize} minimization package's Gaussian process optimizer. 
Figure~\ref{fig:optimizer} shows the optimizer's search over possible values of $(\theta,\phi)$ for the maximum likelihood.

\begin{figure}
    \centering
    \includegraphics[width=.45\textwidth]{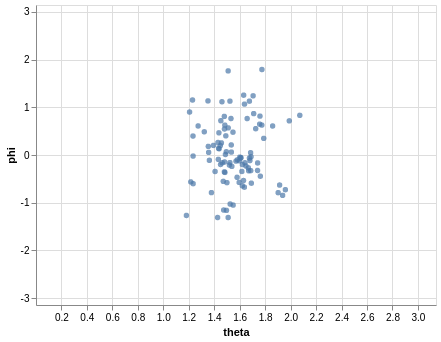}\\
    \includegraphics[width=.45\textwidth]{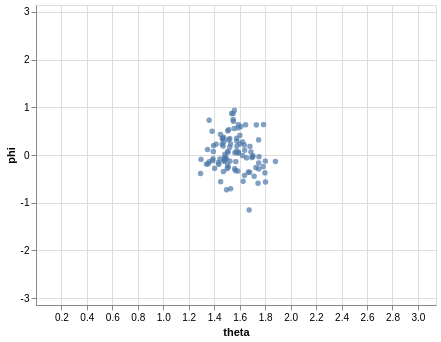}
    \caption{ Maximum-likelihood directions from 100 test ensembles of either 10 (top) or 50 (bottom) neutrons. Using additional neutrons produces a tighter distribution.}
    \label{fig:direction_scatter}
\end{figure}

Ensembles sizes of 10, 50, 250, and 1000 neutrons passing cuts were tested with \num{1000} replications each. 
Figure~\ref{fig:direction_scatter} illustrates the effect of ensemble size on the distribution of the analyzed $(\theta,\phi)$, where the larger ensembles produced tighter distribution of results.
The distribution of output for each ensemble size was fit with a Rayleigh distribution and the width parameter $\sigma$ of that distribution is reported in Table~\ref{tab:pointing_results}.

Table~\ref{tab:pointing_results} also includes a measurement time equivalent to specific ensemble size. 
This is calculated for a scenario with a $10^5$ neutrons/second source \SI{10}{\m} from the array and incorporates the detection efficiency of 12.6\%. 
Measurement time is an important parameter of interest when comparing applications of neutron directionality with different efficiencies. 


\begin{table}
    \centering
    \begin{tabular}{r l l}
    \hline
    $N$, ensemble Size  & Measurement time  & Angular resolution \\
        \hline
        10 & 6.1 s & \SI{25}{\degree} \\
        50 & 31 s & \SI{14}{\degree} \\
        250 & 2.5 min & \SI{8.3}{\degree} \\
        1000 & 10 min & \SI{5.9}{\degree} \\
    \hline
    \end{tabular}
    \caption{Neutron pointing accuracy with different signal statistics.}
    \label{tab:pointing_results}
\end{table}


One limitation of this analysis is that the KDE stores information about the proton recoil spectrum and that information informs the maximum likelihood. 
As such, the KDE cannot be applied to locate a source with an unknown neutron spectrum. 
A future version of this analysis could implement KDEs for a variety of sources and report a joint maximum-likelihood direction and source type.

Another limitation of this study is that gamma backgrounds were not simulated. 
Gammas are expected to very rarely produce single-colored signals due to their long and crooked (detour factor of $\sim 2$) recoil tracks. 
As such, they will add a background in the two-colored region of the KDE. 
This will likely increase the required number of neutrons to achieve a certain precision, thus longer time of detection.
An evaluation of the magnitude of this effect is planned for future work.

\subsection{Anisotropic AMSS for Neutron Spectroscopy}\label{sec:APID_spec}
The sensitivity of the APID AMSS to proton recoil angle can also be applied to neutron spectroscopy measurements. 
The relative angle between the proton recoil and the incident neutron relates the proton recoil energy (easily measured in a scintillator) to the incident neutron energy (often the target of interest in neutron source measurements). 
Although the APID AMSS does not provide precise information on the recoil angle event-by-event, its angular sensitivity disambiguates the thorny problem of neutron unfolding, the reconstruction of the incident neutron spectrum from the recoil spectrum.

The response matrix of a neutron detector describes the distribution of observed proton recoil energies caused by neutrons of given incident energies.
Due to a combination of neutron-proton scattering kinematics, including scattering angles and proton recoil energy, and instrumental uncertainties in scintillation detectors, this response matrix is ill-conditioned. 
Neutron unfolding requires the inversion of this ill-conditioned matrix, generally leading to large systematic errors. 
These errors can be suppressed using various regularization techniques, at the cost of increased errors when a source breaks the assumptions underlying the regularization. 
The APID AMSS adds an additional dimension to this matrix inversion problem, the color fraction measurement. 
This angle-sensitive measurement disambiguates between full-energy proton recoils from lower-energy neutrons and oblique recoils from higher-energy neutrons which would otherwise have identical apparent energies.

The single-source-location simulation set described in Section~\ref{sec:APID} was reused to evaluate the APID AMSS's spectroscopy performance. 
For comparison, a conventional scintillator detector without directional sensitivity was also analyzed. 
This analysis for the conventional detector used the same simulated events without the color fraction, and so had larger photon statistics leading to superior recoil energy resolution. 
The conventional events were used to construct a response matrix with 15 bins of detected photons versus 40 bins of incident energy. 
The APID AMSS results were used to construct a response matrix with four dimensions: 15 bins of detected photons, 10 bins of color fraction, 8 bins of array module number, and 40 bins of incident neutron energy. This response matrix was then flattened into a \num{1200x40} 2D matrix with each row corresponding to a particular combination of the three observables. 
Each matrix was filled with the same number of total events, and so individual bins in the conventional matrix had larger statistics than those from the APID AMSS.

The response matrices were inverted using \texttt{PyUnfold}, a off-the-shelf package implementing an iterative Bayesian approach to spectrum unfolding~\cite{bourbeauPyUnfoldPythonPackage2018}. 
A uniform prior was used, and no additional regularization features were used to evaluate the inherent benefit of the additional APID AMSS observables.

\begin{figure*}
    \centering
    \includegraphics[width=.45\textwidth]{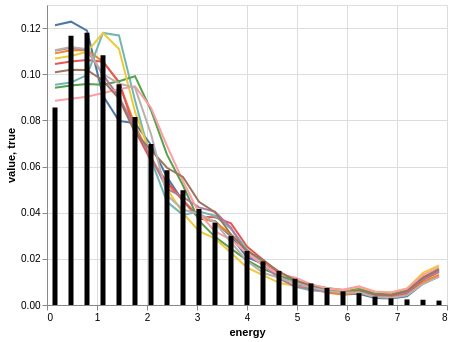}
    \includegraphics[width=.45\textwidth]{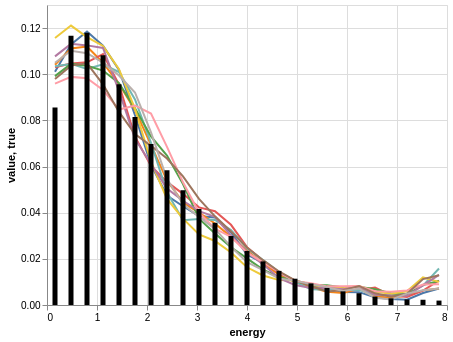}
    \caption{ Unfolding results using a conventional (left) or APID (right) detector. Colored lines represent the unfolded spectra from 10 datasets of 2000 neutrons each. Black bars show the true \ce{^{252}Cf} spectrum.}
    \label{fig:unfolding_specs}
\end{figure*}

The unfolding was tested against 100 ensembles of 2000 detected neutrons each. The results from ten of these ensembles are shown in Figure~\ref{fig:unfolding_specs}. 
As is typical for neutron unfolding, systematic errors appear at the ends of the spectrum where the result is less well constrained by the input. 
To quantify the improvement seen in the APID AMSS analysis, the unfolded output is compared to the true spectrum using a spectral angle mapper (SAM), following \cite{zhuHierarchicalBayesianApproach2019}. 
Over the 100 ensembles, the conventional approach had a mean SAM of $7.7 \pm 0.1$ while the APID AMSS approach had a better mean SAM of $6.7\pm 0.1$.

%% file: narrative/PRMMSS.tex
\section{Position Resolving AMSS}\label{sec:PRMMSS}

Using a gradient  mixture of scintillator dyes, an AMSS can measure the position of a particle's interaction in the scintillator volume. 
This approach can be constructed through spatially varying the ratio of blue and green dye mixture, rather than the zoned dyes discussed in Section \ref{sec:PIDMMSS}. 
This variety of AMSS is referred to here as a Position Resolving (PR) AMSS.


Using two blue-pass or green-pass filtered sensors, the ratio of detected colors depends on the unique mixture of dyes at the interaction position within the PR AMSS scintillator along one dimensional gradient. 
A more complex design uses a conventional position reconstruction to obtain a rough position and employ a periodic gradient to refine that resolution. 
Both designs were evaluated using simulations to demonstrate the achievable position resolution.

The performance of the PR AMSS is compared to the technique used for conventional scintillators in which the signals in two sensors, one at either end of the scintillator bar, are compared to determine interaction position \cite{weinfurtherModelbasedDesignEvaluation2018,liPrototypeSANDDHighlysegmented2019}. 
The distance between the interaction position and each sensor results in differences in the signal amplitude, arrival time, and pulse shape that can be analyzed to reconstruct the position. We refer to this type of reconstruction as ``end-differential'' reconstruction.

\subsection{Position Reconstruction with A Single Gradient}\label{sec:single}

\begin{figure}
    \centering
    \includegraphics[trim={0 5cm 0 4cm}, clip, width=0.48\textwidth]{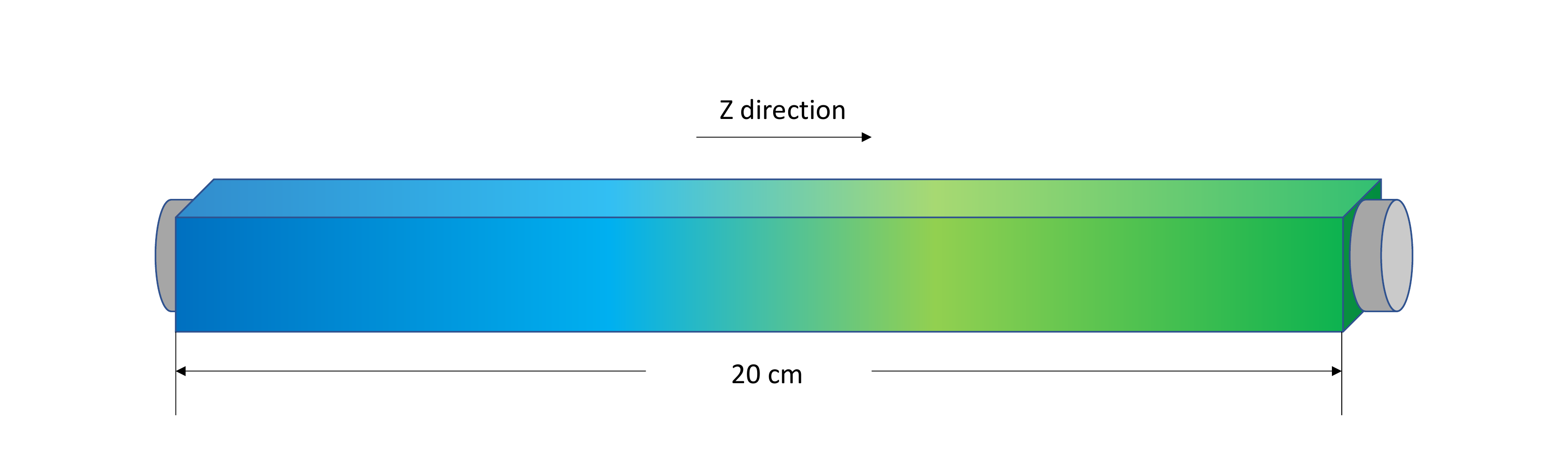}
    \caption{A schematic of the simulated single gradient PR detector. The color shown indicates the color of scintillation light, not the color of scintillator.}
    \label{fig:pr_geom}
\end{figure}

The single-gradient PR detector design was simulated as \SI{1x1x20}{\cm} longitudinal cuboid scintillator with a gradient mixture of blue- and green- emitting dyes varying along its long axis, as illustrated in Figure~\ref{fig:pr_geom}.
This size is identical to that modelled for neutron scatter cameras in \cite{weinfurtherModelbasedDesignEvaluation2018} and similar to the SANDD neutrino detector prototypes~\cite{liPrototypeSANDDHighlysegmented2019}.
For the purpose of light transport simulation, the simulated scintillator was wrapped in a  purely-specular reflector.
Each end of the target was coupled to a filtered photosensor with a round detecting surface. 
To ensure high photon detection efficiency, the green- and blue-filtered photosensors are coupled, respectively, to the green- and  blue-dye-dominated ends of the scintillator.

The behavior of the gradient dye mixture was simulated with a customization to the GEANT4 scintillation process. 
This customized code produces each scintillation photon first by randomly choosing between the green dye and the blue dye according to the relative concentration of those dyes based on the location of energy deposition and then by randomly selecting a photon energy from the emission distribution for the chosen dye. 
At one end of the scintillator ($z=-10$~cm) the dye fraction is 0\% blue, 100\% green; in the center ($z=0$) the dye fraction is 50\% blue, 50\% green; and at the other end ($z=+10$~cm) the dye fraction is 100\% blue. 
The energy response and photon production rate is the same as described in Section~\ref{sec:MC}.
In this simulation, the gradient of dye fraction is continuous, although a realization of this design via AM might yield discrete (but can be optimised to the scale of 50\um) steps.

To evaluate the position reconstruction performance, $5\times10^8$ neutrons with energy uniformly distributed in 0 to 10 MeV were generated from a point source 1 meter away from the scintillator.
Neutrons were emitted isotropically, and given the 1 m standoff arrive approximately uniformly distributed along the $z$-direction of the scintillator volume.
Neutrons were chosen as the simulated particles due to their short interaction tracks so that the position of interaction is well defined. 
For purposes of evaluating the resolution of position reconstruction, the ``true'' interaction position of each photon was defined as the end position of the \texttt{G4Step} with the largest energy deposition.

\begin{figure}
    \centering
    \includegraphics[width=0.45\textwidth]{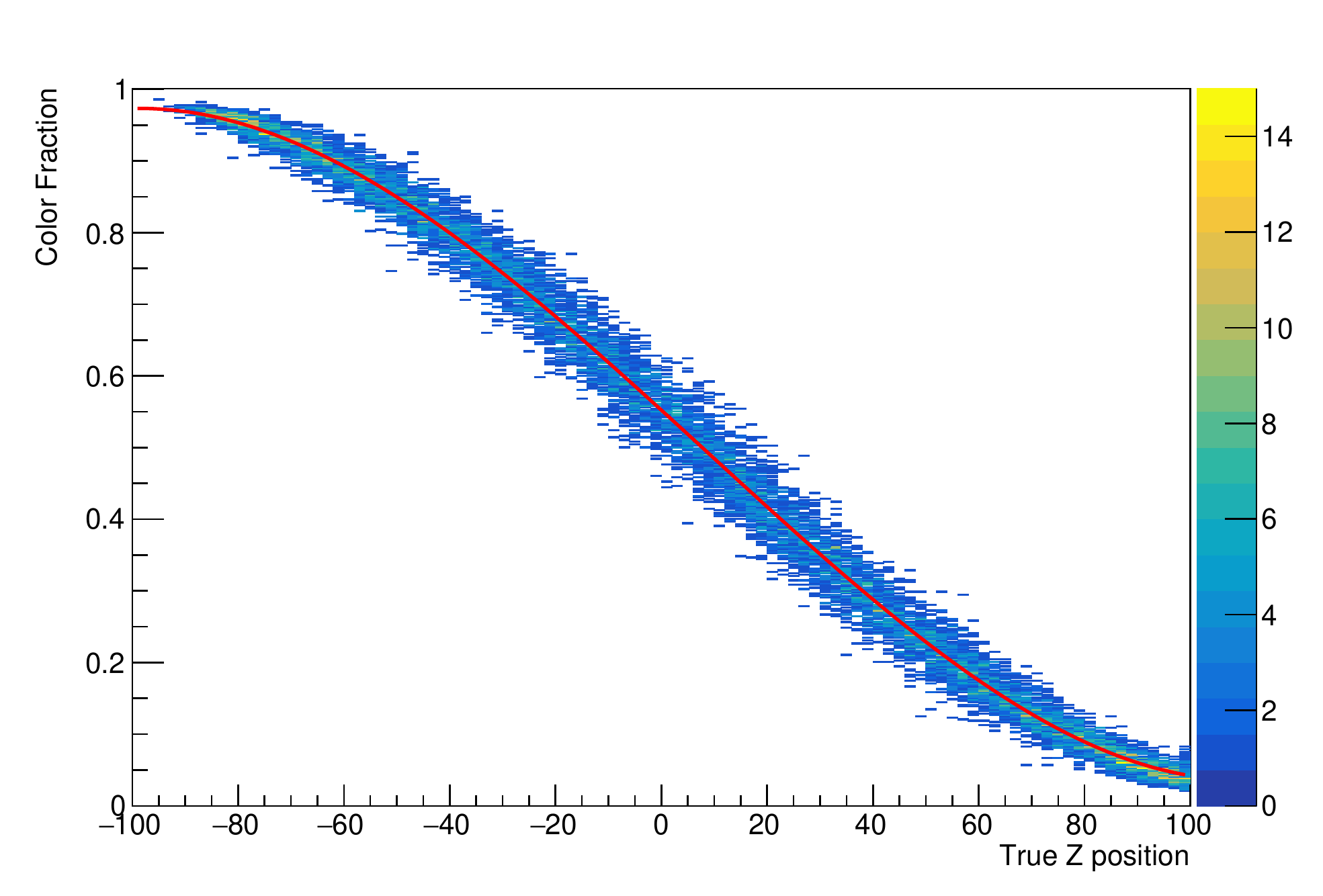}
    \caption{The distribution of neutron-recoil events on green color fractions and their true z-position.
    The distribution is fitted with an 4th degree polynomial function.
    The best fit function is used to reconstruct event position with its color fraction.}
    \label{fig:pr_reco}
\end{figure}

The green color fraction of each event (the fraction of photons with green color to the total number of detected photons) was measured to reconstruct the the positions of neutron interactions. 
A threshold of at least 10 photons detected in each photosensor was applied, resulting in many events at low interaction energy being rejected from the analysis, especially events at either end of the scintillator and thus far from the photosensor at the opposite end.
Figure~\ref{fig:pr_reco} shows the distribution of green color fraction versus true position of the detected interactions across all simulated energies.
This distribution was fitted with a 4th-degree polynomial function to relate color fraction to interaction position.
Notably, although the dye fraction varies linearly, the detected color fraction does not as it is affected by the light attenuation.
The fitted functions was then used to reconstruct event positions using color fraction.

\begin{figure}
    \centering
    \includegraphics[width=0.45\textwidth]{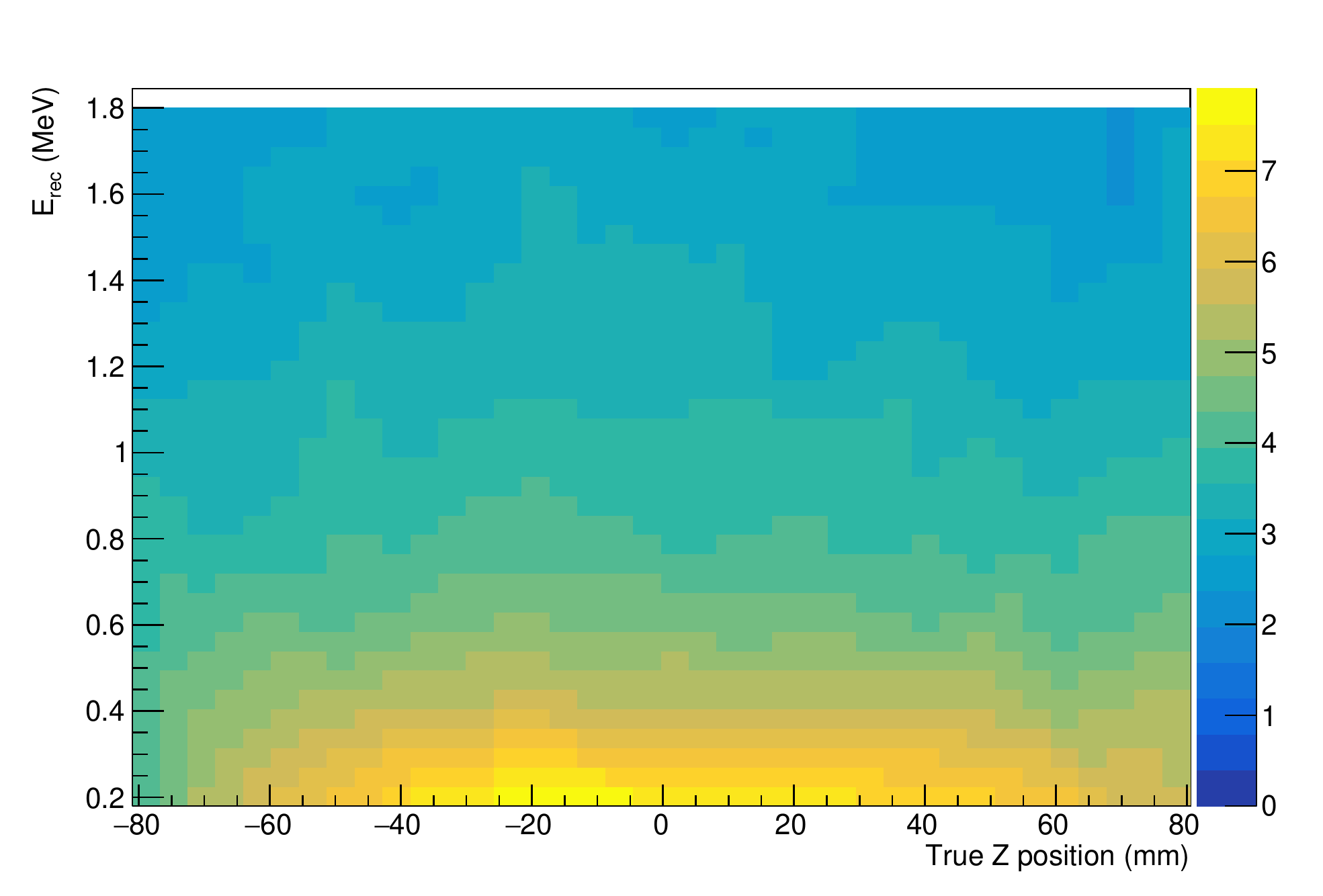}
    \caption{The position resolution using single gradient is dependent on the visible energy and true z-position. Because of the 10-photon thresholds in both sensors, the position reconstruction near the ends of the detector is limited. 
    }
    \label{fig:pr_res}
\end{figure}

The reconstruction position resolution is plotted against the visible energy, as well as the true position of the neutron recoils in the single gradient PR AMSS detector shown in Figure~\ref{fig:pr_res}. 
For events falling in each energy-position bin in figure~\ref{fig:pr_res}, the difference between true event positions and reconstructed events was fitted to a Gaussian distribution with the standard deviation of the best fit Gaussian taken as the position resolution for that bin.
The resolution is better at higher energies and toward the ends of the detector due to more total photons statistics. 
Table~\ref{tab:singleres} lists the resolution results seen in figure~\ref{fig:pr_res} for a few energies of interest.

This position resolution can be compared to that of conventional, end-differential, scintillator detectors.
In reference~\cite{weinfurtherModelbasedDesignEvaluation2018} the measured position resolution using end-differential is 9~mm for events with 1 MeV visible energy. 
The PR AMSS simulated here improves over that resolution by more than a factor of 2. 
The single gradient design that achieves this performance also does not impose any extraodinary timing resolution requirements on the photosensors or scintillator as required in \cite{weinfurtherModelbasedDesignEvaluation2018}. 
However, the use of filtered light sensors does reduce the energy resolution.

\begin{table*}[htb]
    \centering
    \noindent
    \begin{tabular}{r p{30mm} p{30mm}}
    \hline
    Proton recoil energy [MeV]  & Best resolution \ (detector end) (mm)  & Worst resolution (detector center) (mm)\\
        \hline
        \hline
        0.1 & 7.1 & 9.6 \\
        1 & 4.5 & 4.8 \\
        2 & 2.8 & 3.5 \\
        
    \hline
    \end{tabular}
    \caption{Position resolution results for the single gradient design.}
    \label{tab:singleres}
\end{table*}

\subsection{Position Reconstruction with Periodic Gradient Dyed Volumes}

\begin{figure}
    \centering
    \includegraphics[trim={0 5cm 0 4cm}, clip, width=0.45\textwidth]{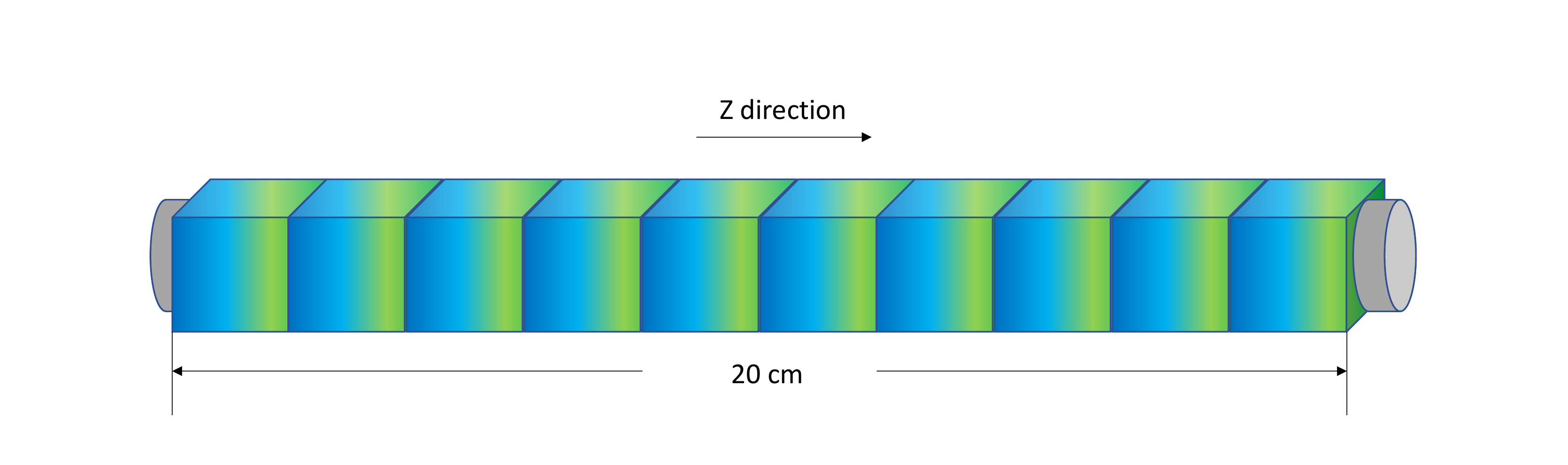}
    \caption{A schematic of the simulated periodic gradient PR detector.}
    \label{fig:ppr_geom}
\end{figure}

The simulation in previous section demonstrated that, by measuring color fraction alone, a single gradient PR AMSS detector can measure event position more precisely than state-of-art end-differential measurements. 
A periodic gradient detector is designed to couple color fraction measurement with end-differential measurement for an even more precise position reconstruction.
In this design, the dye fraction set by periodic sawtooth function, with a period somewhat larger than the end-differential resolution, as illustrated in \ref{fig:ppr_geom}.
The end-differential method provides enough position information to identify which period an interaction occurs in, and then the PR AMSS provides more precise position information within that period.
To employ a metaphor, the end-differential measurement is the hour hand of a clock with lower precision but greater range, and the color fraction is the minute hand that provides more precise resolution once the rough position within the full range is known.

The simulated periodic gradient PR MMSS detector has similar detector geometry to the single gradient PR MMSS simulation, with the only modification being identical periodic gradient dye patterns.
The dimension of the total scintillator volume is 1~cm $\times$ 1~cm $\times$ 20~cm, surrounded by the same reflective films, and coupled to two photon sensors, as shown in figure~\ref{fig:ppr_geom}.
The dye fraction is set by a sawtooth function consisting of ten identical 2~cm periods repeated along the z-axis:
\begin{linenomath}
  \begin{align}
f_{blue}(z) = z/2\textrm{cm} -  floor\left(\frac{z} {2\textrm{cm}}\right) 
  \end{align}
\end{linenomath}    
To be explicit, at $z=-10,-8,-6,...,6,8$~cm, the dye fraction is 100\% green. 
At $z=-9,-7,-5,...,5,7,9$~cm, the dye fraction is 50\% blue, 50\% green. 
And, just before the dye fraction turns back to green, it is 100\% blue (i.e., at $z=-8,-6,...,6,8,10$~cm - $\epsilon$).
As with the single gradient, $5\times10^8$ neutrons were simulated to test the resolution.

\begin{figure}
    \centering
    \includegraphics[width=0.45\textwidth]{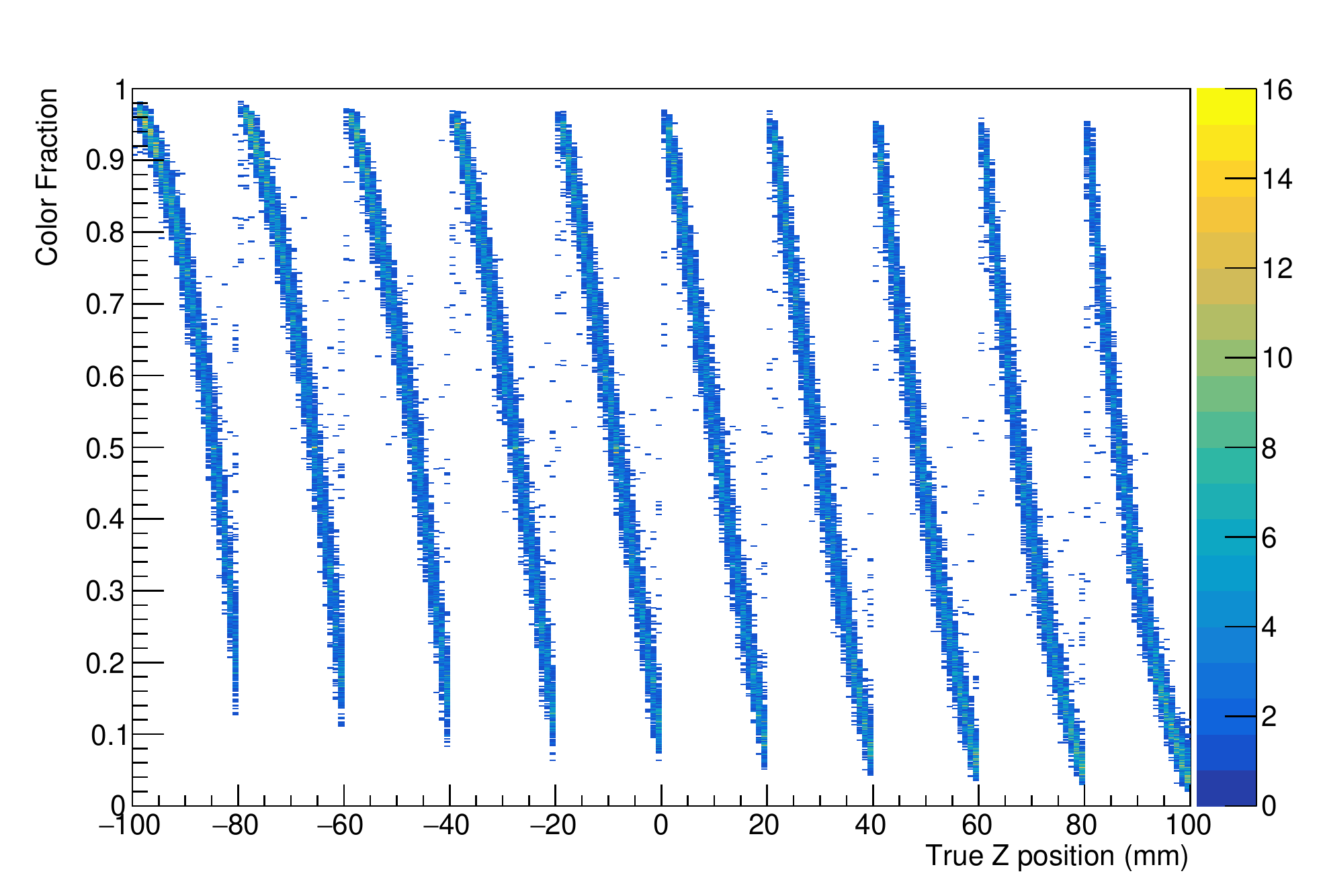}
    \caption{Color fractions distributed along the true z-positions.
    The non-uniform distribution among periods caused by the threshold effect and photon attenuation.}
    \label{fig:ppr_reco}
\end{figure}

The observed green color fraction varies from 1 to 0, from one end to the other of each period, as shown in figure~\ref{fig:ppr_reco}. 
As in the single gradient case, the linear variation of the dye within each period translates into a nonlinear variation in measured color fraction due to the simulated light transport efficiency.

Rather than reproduce the simulation and analysis of the end-differential effects in reference~\cite{weinfurtherModelbasedDesignEvaluation2018}, this study references Figure~29 from that reference to emulate the position resolution available as a function of energy. 
For each event, an end-differential position measurement was assigned by adding a Gaussian random value to the true position with the standard deviation of that Gaussian interpolated as a function of energy from Figure~29 in \cite{weinfurtherModelbasedDesignEvaluation2018}. 
The loss of light in the AMSS's color-filtered sensors was accounted for when identifying the end-differential resolution for a given observed energy.

\begin{figure}
    \centering
    \includegraphics[width=0.48\textwidth]{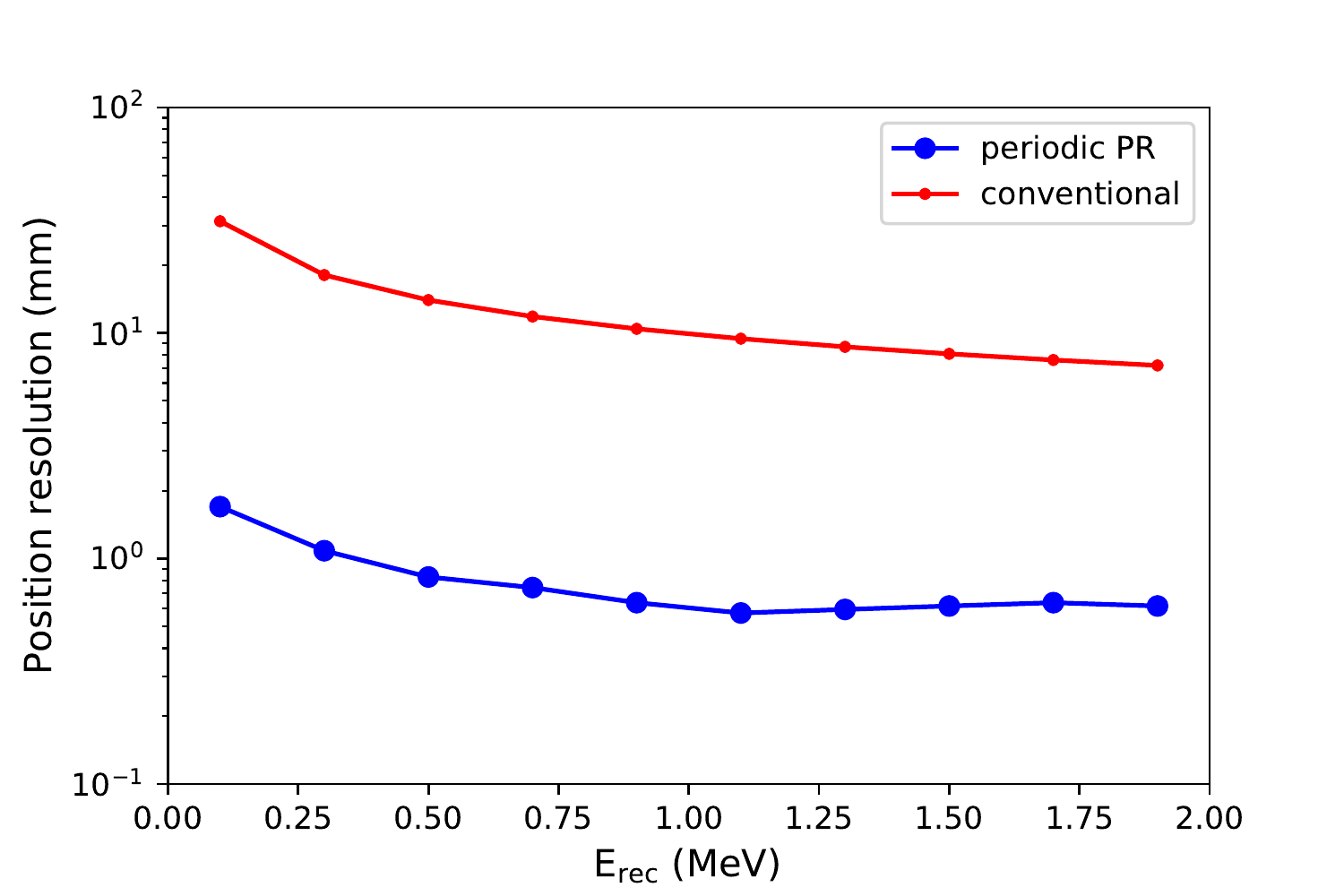}
    \caption{Position resolution of the periodic gradient PR AMSS, compared with the end-differential position reconstruction, as a function of visible energy.}
    \label{fig:ppr_res}
\end{figure}

The color fraction and end-differential position measurement were used as two inputs of a  support-vector regression (SVR) model to infer the final reconstructed position. 
The SVR enables an optimized analysis of these two correlated variables, including the effects of the nonlinear color fraction curve and the edges of the sawtooth function. 
As before, the resolution of this technique is quantified as the standard deviation of a Gaussian fit to the distribution of the difference between the the final reconstructed z-position and the true z-position.
This distribution of this difference is calculated for various visible energy and the analyzed resolution is shown in Figure~\ref{fig:ppr_res}. 
The end-differential resolution by itself (without the ``penalty'' from filtered sensors) is shown as well to illustrate the improvement resulting from the periodic gradient. 
The resolution of the periodic gradient PR AMSS detector is better than a purely end-differential detector by an order of magnitude across all energies. For interactions with energy > 0.5 MeV, the position resolution reaches sub-millimeter range in the 20~cm longitudinal detector. 


The effectiveness of the periodic gradient technique is sensitive to both the period size and the resolution of the end-differential reconstruction. 
If the end-differential reconstruction fluctuates by more than one period from the true value then the final reconstructed position will also be an entire period off. 
A small number of period sizes have been tested in simulation thus far, which have demonstrated that reducing the period size improves the position resolution up to a point, after which the period is small enough that the end-differential reconstruction produces a whole-period error.